\documentclass[twocolumn,aps,prd,superscriptaddress,preprintnumbers,amsmath,amssymb,tightenlines,longbibliography,nofootinbib]{revtex4-1}
\usepackage{epsfig}  
\usepackage{graphicx}
\usepackage{color}
\usepackage[dvipsnames]{xcolor}
\usepackage{float}
\usepackage{amsfonts}
\usepackage{amsmath}
\DeclareMathOperator{\sign}{sgn}
\usepackage[compat=1.0.0]{tikz-feynman}
\usepackage{slashed}
\usepackage{verbatim}
\usepackage{mathrsfs}
\usepackage{slashed}
\usepackage{soul}
\usepackage{braket}
\usepackage{dsfont}
\usepackage[colorlinks,citecolor=blue,urlcolor=blue,bookmarks=false,hypertexnames=true]{hyperref}
\usepackage{orcidlink}

\large

\begin{document} 

\title{Role of the GPD-$E$ in exclusive dijet electroproduction at the Electron-Ion Collider}

\author{Trambak Jyoti Chall\orcidlink{000000025932278X}}
\email{trambakjyotic@dokt.ur.edu.pl}
\affiliation{Doctoral School, Institute of Physical Sciences, University of Rzeszów, Al. Tadeusza Rejtana 16C, PL-35-959 Rzeszów, Poland}

\author{Agnieszka Łuszczak\orcidlink{0000000179685388}}
\email{agnieszka.luszczak@pk.edu.pl}
\affiliation{T. Kosciuszko Cracow University of Technology, PL-30-084 Kraków, Poland}

\author{Marta Łuszczak\orcidlink{0000000317768019}}
\email{mluszczak@ur.edu.pl}
\affiliation{Institute of Physical Sciences, University of Rzeszów, Al. Tadeusza Rejtana 16C, PL-35-959 Rzeszów, Poland}

\author{Wolfgang Schäfer\orcidlink{0000000207640372}}
\email{wolfgang.schafer@ifj.edu.pl}
\affiliation{The Henryk Niewodniczański Institute of Nuclear Physics, Polish Academy of Sciences, ul. Radzikowskiego 152, PL-31-342 Kraków, Poland}

\author{Antoni Szczurek\orcidlink{0000000152478442}}
\email{antoni.szczurek@ifj.edu.pl}
\affiliation{Institute of Physical Sciences, University of Rzeszów, Al. Tadeusza Rejtana 16C, PL-35-959 Rzeszów, Poland}
\affiliation{The Henryk Niewodniczański Institute of Nuclear Physics, Polish Academy of Sciences, ul. Radzikowskiego 152, PL-31-342 Kraków, Poland}

\begin{abstract}
{
We calculate the contribution of the $E$-type generalized parton distributions (GPDs) to exclusive dijet electroproduction in $ep\to e'jjp$ in collinear QCD factorization, including both the proton helicity-flipping amplitude and the GPD-$E$ term in the helicity-conserving amplitude, using the double distribution approach for modeling. We include light sea and valence-quark exchanges as well as gluon exchanges and compare the proton helicity-conserving and helicity-flipping contributions for HERA-energy phase space and for Electron-Ion Collider (EIC) kinematics. We find that the helicity-flipping cross section is dominated by valence-quark exchange and has a markedly different dependence on $|t|$ and $x_{\mathbb P}$ from the helicity-conserving contribution. In the EIC kinematics considered here, the helicity-flip contribution becomes comparable to the helicity-conserving one at large $|t|$ and $x_{\mathbb P}$. The results show that exclusive dijet electroproduction can be sensitive to the $E$-type GPDs in the kinematic region accessible at the EIC.
}
\end{abstract}

\maketitle
\newpage

\section{Introduction}
\label{sec:1}
Exclusive dijet production in diffractive deep inelastic scattering is a useful process for studying the partonic structure of the proton. In the collinear factorization approach, generalized parton distributions (GPDs)  enter as the nonperturbative hadronic input. For an unpolarized spin-$\frac{1}{2}$ proton, the leading-twist chiral-even sector is described by the GPDs $H^{g,q}$ and $E^{g,q}$~\cite{Diehl:2003ny,Belitsky:2005qn,Ji:1998pc}.
In the forward limit, $H^{g,q}$ is related to the usual unpolarized parton distributions, whereas there is no analogous forward parton density that determines $E^{g,q}$. For valence quarks, the first moment of $E^q$ is constrained by the Pauli form factor. The $E$-type GPDs are therefore considerably less known than the corresponding $H$-type distributions. For recent reviews on GPDs, see~\cite{Liuti:2026jid,Guo:2026swz}.

In our previous work~\cite{Chall:2026oes}, we studied exclusive dijet electroproduction in $ep\to e'jjp$ in collinear QCD factorization, based on the approach developed in~\cite{Braun:2005rg} including gluon, sea-quark and valence-quark exchanges. For a similar study, see~\cite{Pang:2026lsr}. We considered the proton helicity-conserving contribution and presented results for a broad region of phase space, including the HERA kinematic region. We found that the valence contribution, although relatively small at small $x_{\mathbb P}$, has a different shape from the gluon and sea contributions and becomes increasingly important for larger $x_{\mathbb P}$.
The analyses of~\cite{Chall:2026oes,Pang:2026lsr} were restricted to the proton helicity-conserving part. At small $x_{\mathbb P}$, and consequently at small skewness $\xi$, the $E^{g,q}$ term in the helicity-conserving amplitude is suppressed and the contribution from $H^{g,q}$ dominates. In the present work we include the proton helicity-flip amplitude induced by the Pauli coupling, which is directly proportional to $E^{g,q}$, and investigate its importance outside the small-$x_{\mathbb P}$ region. 
We stress that ``helicity flip'' in this work refers to the proton transition $\lambda'\neq\lambda$. The helicity of the active parton is conserved and we work with the chiral-even GPDs $H^{g,q}$ and $E^{g,q}$. We do not consider in this work parton-helicity-flip, nor transversity GPDs.

The EIC~\cite{Aschenauer:2017jsk} will extend measurements of hard exclusive reactions to a broad kinematic region and different center-of-mass energies~\cite{AbdulKhalek:2021gbh}. Recently, exclusive quark and gluon dijet production at collider energies has also been discussed in the collinear GPD framework, including predictions for EIC kinematics. In the numerical analysis of~\cite{Chall:2026oes,Pang:2026lsr}, the $E^{g,q}$ contributions were neglected with respect to the dominant $H^{g,q}$ terms. Here we concentrate specifically on the proton helicity-flip contribution generated by $E^{g,q}$.
We find that the helicity-flip contribution becomes increasingly important toward larger $|t|$ and larger $x_{\mathbb P}$. The aforementioned contribution is also seen to be dominating over the gluon and sea-quark exchange contributions for EIC kinematics.
This makes the moderate- and large-$x_{\mathbb P}$ region particularly interesting for studying the $E$-type GPDs.

The paper is organized as follows. In Sec.~\ref{sec:2} we summarize the theoretical formalism and give the proton helicity amplitudes. In Sec.~\ref{sec:3} we specify the GPD input used in the calculation, with particular emphasis on the $E$-type GPDs. In Sec.~\ref{sec:4} we present the results for the HERA-energy reference calculation and for EIC kinematics. The conclusions are given in Sec.~\ref{sec:5}.

\section{Theoretical Formalism}
\label{sec:2}
We consider the exclusive electroproduction of a quark--antiquark dijet pair,
\begin{equation}
e(l)+p(p)\to e(l')+q(p_1)+\bar q(p_2)+p(p')\,,
\end{equation}
with virtual-photon momentum $q=l-l'$ and virtuality $Q^2=-q^2$. The momentum transferred to the proton is $\Delta=p'-p$, with $\Delta^2<0$, and the average proton momentum is $P=(p+p')/2$. The invariant mass squared of the diffractively produced dijet system is
\begin{equation}
M^2=(q-\Delta)^2\,.
\end{equation}
The asymmetry parameter can be expressed as,
\begin{equation}
\xi=\frac{(p-p')\cdot n_-}{(p+p')\cdot n_-}\,.
\end{equation}
As in~\cite{Chall:2026oes}, the diffractive momentum fraction can be expressed as,\begin{equation}
x_{\mathbb P}
=
\frac{M^2+Q^2-t}{W^2+Q^2}
\simeq
\frac{M^2+Q^2}{W^2+Q^2}\,,
\qquad
\frac{2\xi}{1+\xi} = x_{\mathbb P}\,,
\label{xp-xi}
\end{equation}
where the approximate relations are used in the high-energy kinematics considered here.

The chiral-even quark and gluon GPDs relevant for an unpolarized spin-$\frac{1}{2}$ proton enter through the standard Lorentz decomposition:
\begin{widetext}

\begin{equation}
F^{g,q}\left(x,\xi,\Delta^2\right)=\frac{1}{2(P\cdot n_-)}
\left[
H^{g,q}(x,\xi,\Delta^2)\,
\bar u(p')\slashed n_-u(p)
+
E^{g,q}(x,\xi,\Delta^2)\,
\bar u(p')\frac{i\sigma^{\alpha\beta}n_{-\alpha}\Delta_\beta}{2m_N}u(p)
\right]\,,
\label{HE}
\end{equation}

\end{widetext}
where $m_N$ is the nucleon mass. Projecting this matrix element onto light-cone proton helicity states gives~\cite{Diehl:2003ny,Diehl:2001pm}
\begin{eqnarray}
    F^{g,q}_{+,+}&=&F^{g,q}_{-,-}=\sqrt{1-\xi^2}\left[H^{g,q}-\frac{\xi^2}{1-\xi^2}E^{g,q}\right]\,,\nonumber\\
    F^{g,q}_{-,+}&=&-\left[F^{g,q}_{+,-}\right]^{\ast}=\sqrt{1-\xi^2}\left[\frac{|\vec{D}|e^{i\varphi}}{2m_{N}}E^{g,q}\right]\,,
    \label{str}
\end{eqnarray}
where we see that the GPD-$E$ contributes to both the helicity conserving and flipping amplitudes and consequently, the differential cross sections.
From Eq.~\eqref{str} we see that at small skewness the proton helicity-conserving amplitude is dominated by $H^{g,q}$ , since the $E^{g,q}$ term is multiplied by $\xi^2/(1-\xi^2)$. On the other hand, the proton helicity-flip amplitude is directly proportional to $E^{g,q}$ and to the momentum-transfer factor $|\vec D|/(2m_N)$. Thus, the smallness of the $E^{g,q}$ term in $F_{+,+}$ does not imply that the $F_{-,+}$ contribution can be neglected. 
The vector $D$ is given by
\begin{equation}
    D=\frac{p^{\prime}}{1-\xi}-\frac{p}{1+\xi}\,,
\end{equation}
and the phase factor appearing in the helicity flipping part arises from the transformation of the light-cone helicity states under a rotation around the $z$-axis by an angle $\varphi$, where $\varphi$ is the azimuthal angle of $\vec{D}$ in the transverse $x-y$ plane measured with respect to the $z$-axis. Such a factor will of course drop out at the level of the cross section due to squaring of such amplitudes.
We know that transverse boosts are Lorentz transformations that leave the plus-component of the vector $D^+$ invariant and the transverse component (in the $x-y$ plane) shifts like $\vec{D}\rightarrow \vec{D}-D^+ \vec{v}$. Now, since by construction $D^{+}=0$, $\vec{D}$ is invariant under such transverse boost parameterized by a two dimensional boost velocity vector $\vec{v}$.
Following the conventions of~\cite{Diehl:2003ny}, the invariant momentum transfer can be written as
\begin{equation}
t=-t_0+(1-\xi^2)\vec D^{\,2}\,,
\qquad
t\geq -t_0=\frac{4\xi^2m_N^2}{1-\xi^2}\,,
\label{t0D}
\end{equation}
where $t$ is  Lorentz invariant under transverse boost and $t_0$ is the value of $t$ at vanishing transverse momentum transfer.

So the pre-factor $\frac{|\vec{D}|}{2m_{N}}$ can be expressed as
\begin{equation}
    \frac{|\vec{D}|}{2m_{N}}=\frac{1}{1-\xi^2}\sqrt{\left(1-\xi^2\right)\left(1+\frac{t}{4m_N^2}\right)-1}
\end{equation}

The phase $e^{i\varphi}$ drops out after taking the modulus squared of the amplitude. The same convention gives
\begin{equation}
t=-\frac{\vec\Delta_\perp^{\,2}+x_{\mathbb P}^2m_N^2}{1-x_{\mathbb P}}\simeq-\frac{\vec\Delta_\perp^{\,2}+4\xi^2m_N^2}{1-\xi^2}\,,
\label{tmand}
\end{equation}
so that the quantity shown as a positive momentum-transfer variable in the numerical plots is $|t|=-t$.

Our dijet cross section, summed over the parton helicities and color of the final-state $q\bar q$ pair, can be written in the standard form, putting into evidence the contributions of transversely and longitudinally polarized photons and their relevant interferences. We retain the normalization and hard-scattering conventions of~\cite{Chall:2026oes},
\begin{widetext}
\begin{eqnarray}
\frac{d\sigma (ep \to e'jj p)}{dy\, dQ^2 d \phi d\vec p_\perp^{~2} d \vec \Delta_\perp^2 dz} &=& \sum_{f} \frac{\alpha_s^2 \alpha_{em}^2 e_f^2}{16 \pi^2 N_c (1- \xi^2)Q^2 y} \frac{1}{(\vec p_\perp^{~2} + m_f^2 + z(1-z) Q^2)^4}\nonumber\\ &\times&
\Bigg[\frac{1 + (1 - y)^2}{2}  \Omega^f_T
+ (1-y) \Omega^f_L - (2-y) \sqrt{1-y} \cos \phi\, \Omega^f_{LT}
- 2(1 - y) \cos 2\phi\, 
\Omega^f_{TT} \Bigg],
\label{cs}
\end{eqnarray}
where $\phi$ denotes the azimuthal angle between the leptonic plane and the dijet plane, $e_f$ and $m_f$ are the fractional electric charge and mass of a quark of flavor $f\in\{{\rm up,\,down,\,strange}\}$, $y$ is the DIS inelasticity, $z$ is the longitudinal momentum fraction carried by the quark jet, and $\vec{p_{\perp}}$ is the jet transverse momentum.
For each quark flavor,
\begin{eqnarray}
    \Omega^f_T &=& m_f^2 |{\cal I}^g_L|^2 + \vec p_\perp^{~2} \Big( |(1-z) {\cal I}^g_T + 2 C_F {\cal I}^{q_1}_T|^2 + |z {\cal I}^g_T - 2 C_F {\cal I}^{q_2}_T|^2 \Big), \nonumber \\
   \Omega^f_L &=& 4 z^2 (1-z)^2 Q^2 \,   |{\cal I}_L^g + 2C_F {\cal I}_L^{q}|^2, 
\end{eqnarray}
and,
\begin{eqnarray}  
   \Omega^f_{LT}  &=& 2 z (1-z) Q \, |\vec p_\perp|\, \, \Re e \left[ \left({\cal I}_L^g + 2C_F {\cal I}_L^{q}\right) \left( (1 - 2z) {\cal I}_T^g + 2C_F ({\cal I}_T^{q_1} + {\cal I}_T^{q_2}) \right)^{\ast} \right], \nonumber \\
   \Omega^f_{TT} &=& \vec p_\perp^{~2} \,  \Re e \left[ \left((1-z) {\cal I}_T^g + 2C_F {\cal I}_T^{q_1}\right)\left(z {\cal I}_T^g - 2C_F {\cal I}_T^{q_2}\right)^{\ast} \right].
   \label{cs2}
\end{eqnarray}

\end{widetext}
Here $C_F=(N_c^2-1)/(2N_c)$  and $N_c=3$. The form-factor integrals ${\cal I}$ contain the convolution of the hard-scattering kernels with the corresponding quark and gluon GPDs; their explicit hard kernels are unchanged from~\cite{Braun:2005rg,Chall:2026oes}.

The essential new ingredient relative to~\cite{Chall:2026oes,Pang:2026lsr} is therefore not a modification of the hard dijet subprocess but the retention of the proton-helicity-flip matrix element proportional to $E^{g,q}$.

For an unpolarized initial proton and with the recoil-proton helicity unobserved, the cross section contains the sum over the final proton helicities. Schematically, using the parity relations for a spin-$\frac{1}{2}$ target,
\begin{equation}
d\sigma_{\rm unpol}\ \propto\
\frac{1}{2}\sum_{\lambda,\lambda'} |F_{\lambda',\lambda}|^2= |F_{+,+}|^2+|F_{-,+}|^2\,.
\label{unpol-helicity-sum}
\end{equation}
Thus, the separate $F_{+,+}$ and $F_{-,+}$ curves shown below represent a decomposition of the calculated unpolarized cross section and are not, by themselves, two separated experimental cross sections. Their comparison is useful for identifying the kinematic region in which the $E$-type GPD contribution becomes important. A direct experimental separation of the two proton helicity amplitudes would require additional polarization information.
The modeling of the GPDs $H^{g,q}$ and $E^{g,q}$ used in the calculation is specified in the next section.
\section{Modeling Helicity-flip GPDs}
\label{sec:3}
For modeling the GPDs, we use the Double Distribution (DD) representation, wherein the polynomiality of the resulting GPDs is warranted by construction.
Since the modeling of the $H$-type GPDs is unchanged with respect to our previous work~\cite{Chall:2026oes}, we follow the same conventions here and summarize only the ingredients needed for the present calculation and give more details for the $E$-type GPDs, which enter explicitly in the proton helicity-flip amplitude.

To explicitly describe the distinct domains in $(x,\xi)$ space for integration over $\beta$, we write for $F=H,E$
\begin{eqnarray}
    F^i\left(x,\xi,\Delta^2\right)&=&\int_{-1}^1 d\beta \frac{1}{\xi}\, g^i_F\left(\beta,\frac{x-\beta}{\xi},\Delta^2\right)\nonumber\\&\times&\Xi\left(\beta|x,\xi\right)\,,
\end{eqnarray}
where $i\in\{{\rm gluon,sea,valence}\}$ and the function $\Xi(\beta|x,\xi)$ divides the $\beta$ region into the allowed DGLAP and ERBL support regions as,
\begin{eqnarray}
    \Xi\left(\beta|x,\xi\right)&=& \theta\left(x>\xi\right)\theta\left(\frac{x+\xi}{1+\xi}>\beta>\frac{x-\xi}{1-\xi}\right) \nonumber\\&+& \theta\left(x<-\xi\right)\theta\left(\frac{x+\xi}{1-\xi}>\beta>\frac{x-\xi}{1+\xi}\right)\nonumber\\&+&\theta\left(-\xi<x<\xi\right)\theta\left(\frac{x+\xi}{1+\xi}>\beta>\frac{x-\xi}{1+\xi}\right)\nonumber\,.\\
\end{eqnarray}
To model the DDs, we use the Radyushkin Ans\"{a}tze with factorized $\Delta^2$ dependence as~\cite{Radyushkin:1998es, Radyushkin:1998bz, Goloskokov:2006hr},
\begin{equation}
g^i_{h,e}\left(\beta,\alpha,\Delta^2\right)=f^i_{h,e}\left(\beta\right)\pi_i\left(\beta,\alpha\right)\, {\cal F}^i_{h,e} (\Delta^2)\,.
\label{DDmodel}
\end{equation}
The profile function $\pi_i\left(\beta,\alpha\right)$ in Eq.~\eqref{DDmodel} introduces the skewness dependence via $\alpha$ and can be chosen as
\begin{equation}
    \pi_i\left(\beta,\alpha\right)=\frac{\Gamma(2n_i + 2)}{2^{2n_i+1} \, \Gamma^2(n_i + 1)} 
\frac{[(1 - |\beta|)^2 - \alpha^2]^{n_i}}{(1 - |\beta|)^{2n_i + 1}}
\label{prof}
\end{equation}
with 
\[
n_i = \left \{\begin{matrix} 1 \quad {\rm for\,\,i=valence} \\ \quad 2\quad {\rm for\,\,i=gluon,\,sea}
\end{matrix}\right.
\]
such that the profiles are normalized as,
\begin{figure*}
    \centering   
    \includegraphics[width=0.43\linewidth]{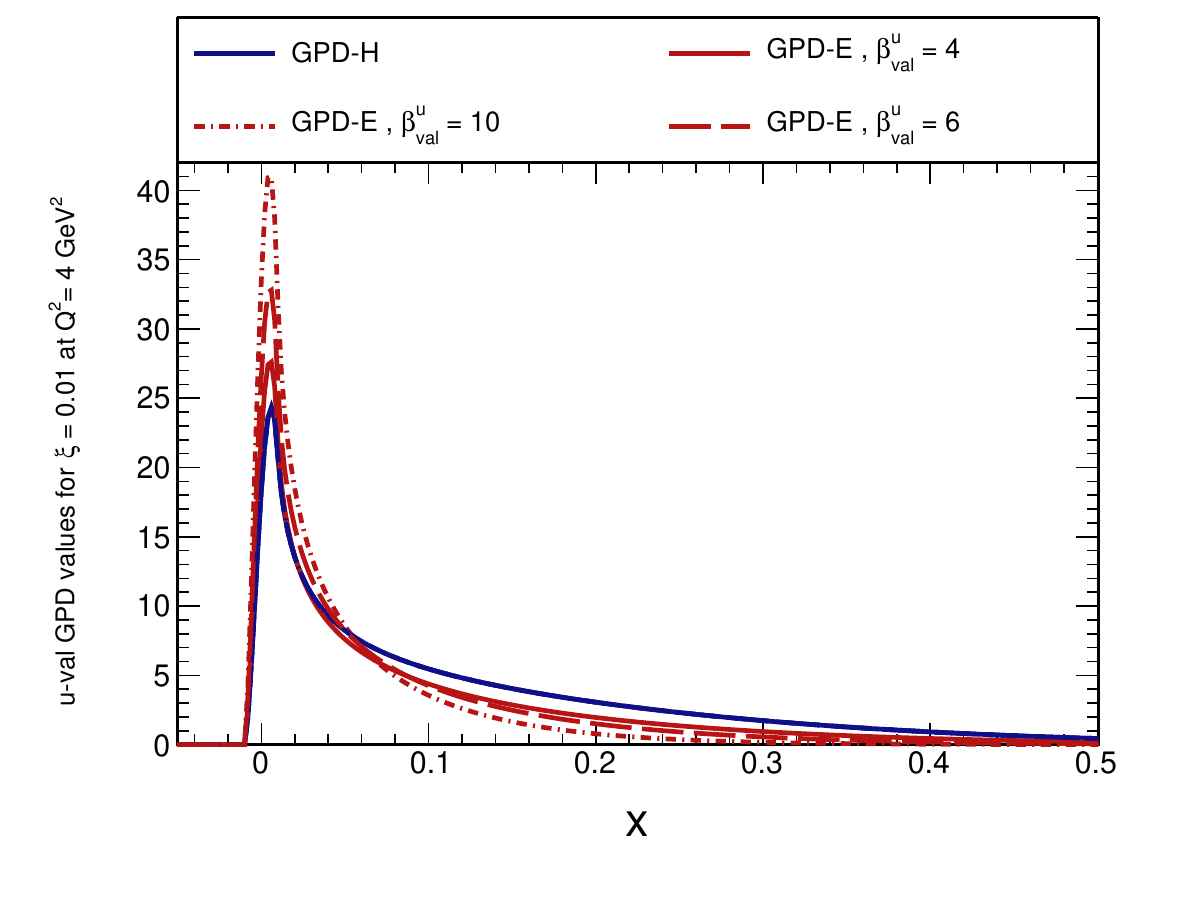} \hspace{1em}
    \includegraphics[width=0.43\linewidth]{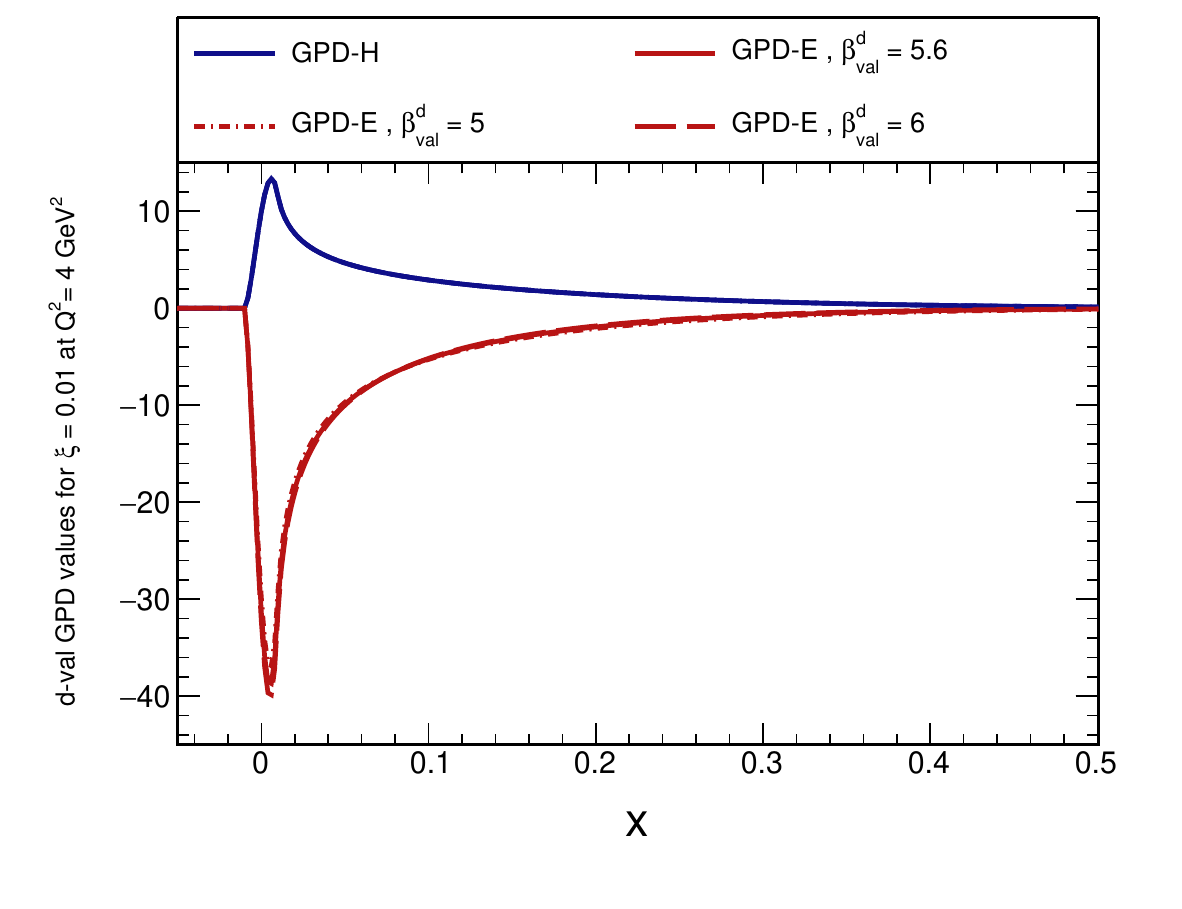} \hspace{1em}
    \includegraphics[width=0.43\linewidth]{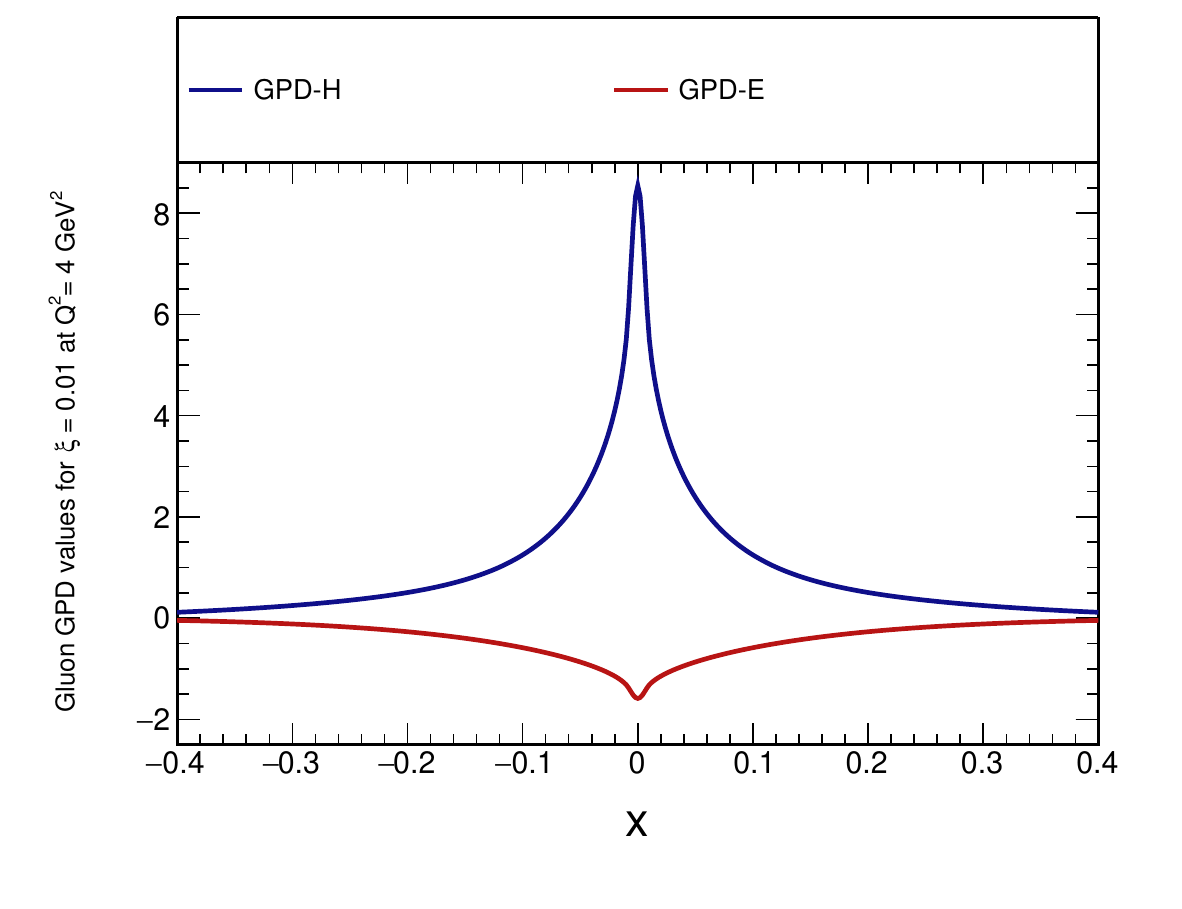} \hspace{1em}
    \includegraphics[width=0.43\linewidth]{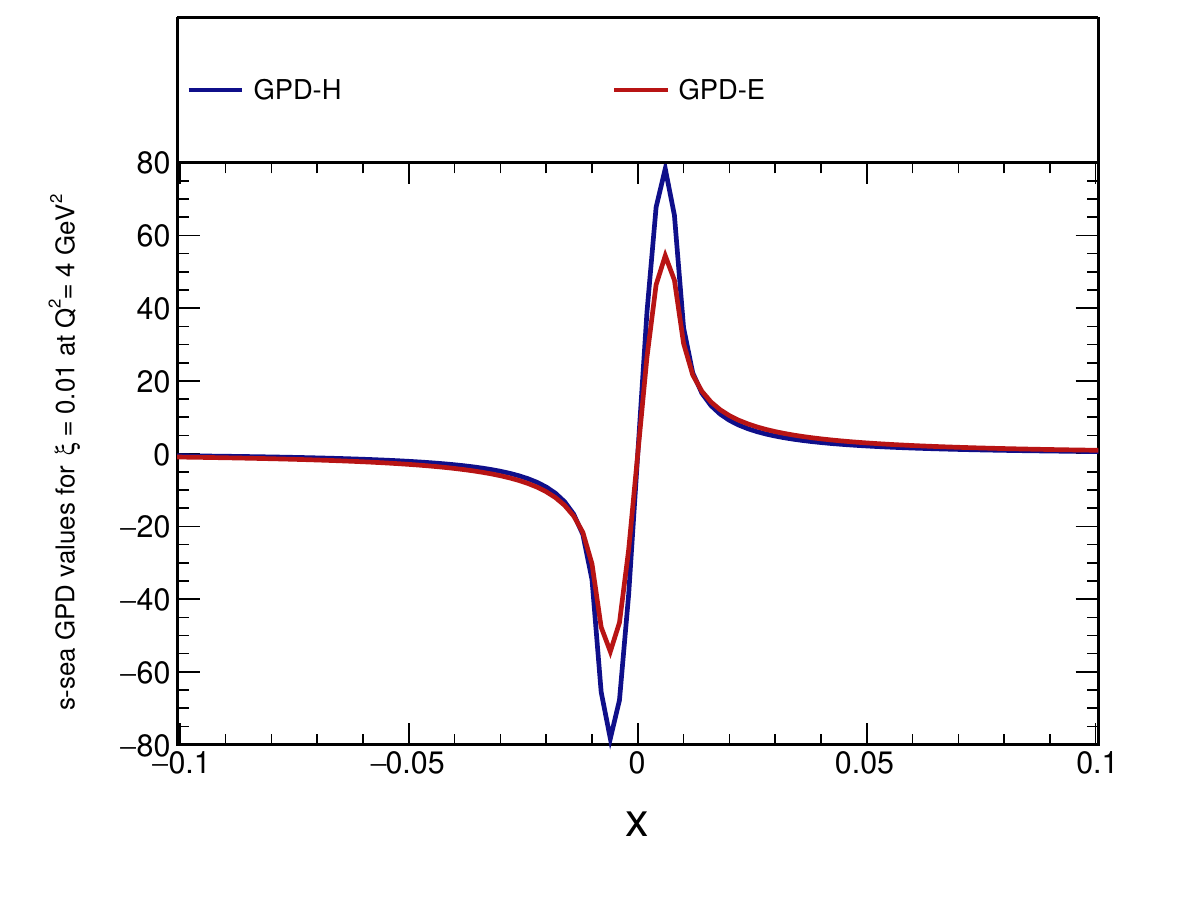}
    \caption{The figures show the H and E-type GPDs used in this work at $\Delta^2=0$ as a function of the momentum fraction $x$ for $\xi=0.01$ and at scale $\mu^2=4\,\rm GeV^2$. The top-left and top-right panels show the $u$ and $d$-valence GPDs, respectively. The bottom-left panel shows the gluon GPDs and the bottom-right panel shows the $s$-sea quark GPDs. The solid blue curves show the GPD-$H$, the solid red curves show the GPD-$E$. In the upper panel the solid red curves correspond to the first variant of the $\beta_{\rm val}^{q=u,d}$, the dash-dotted red curves show the second variant, while the dashed red curves show the third variant corresponding to Table. \ref{tab}.}
    \label{fig:x1}
\end{figure*}
\begin{equation}
\int_{-1+|\beta|}^{1-|\beta|} d\alpha\,\pi_i(\beta,\alpha)=1\,.
\end{equation}
The profile function in Eq.~\eqref{prof} is even in $\alpha$ for all partonic components. The different symmetry properties of the quark and gluon DDs are carried by the corresponding forward-limit functions in $\beta$, not by $\pi_i(\beta,\alpha)$.

For the $H$-type GPDs, the forward-limit functions are taken in the same form as in~\cite{Chall:2026oes} and can be given as,
\begin{eqnarray}
f_h^i(\beta)=
\begin{cases}
|\beta|\,y_i(|\beta|)\,, & i={\rm gluon}\,,\\
y_i(|\beta|)\,\sign(\beta)\,, & i={\rm sea}\,,\\
y_i(\beta)\,\theta(\beta)\,, & i={\rm valence}\,,
\end{cases}
\label{fh}
\end{eqnarray}
where $y_i$ are the usual PDFs corresponding to the exchanged parton. As far as the collinear parton input is concerned, we use the CT18NLO PDF set, as in~\cite{Hou:2019efy} at the scale $\mu^2=4\rm GeV^2$.
\begin{figure*}
    \centering   
    \includegraphics[width=0.43\linewidth]{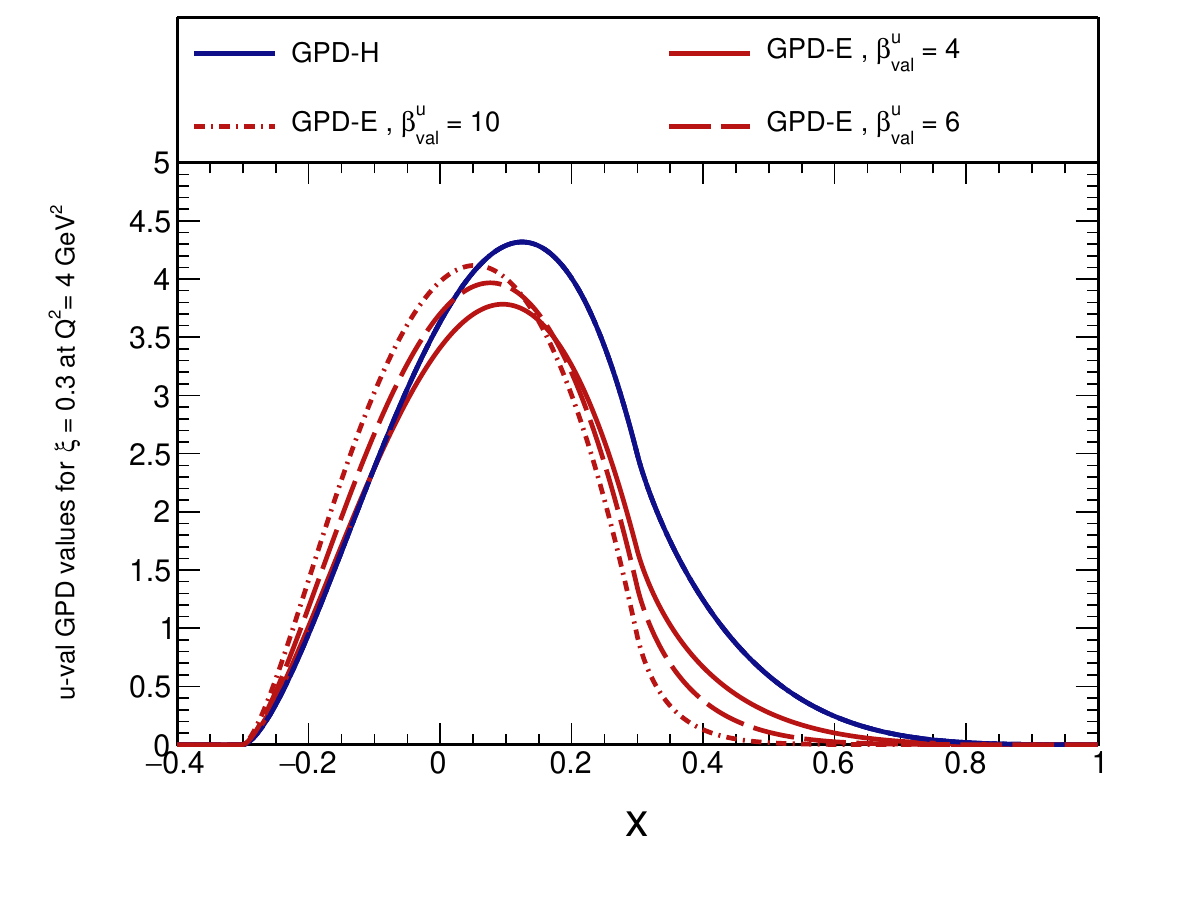} \hspace{1em}
    \includegraphics[width=0.43\linewidth]{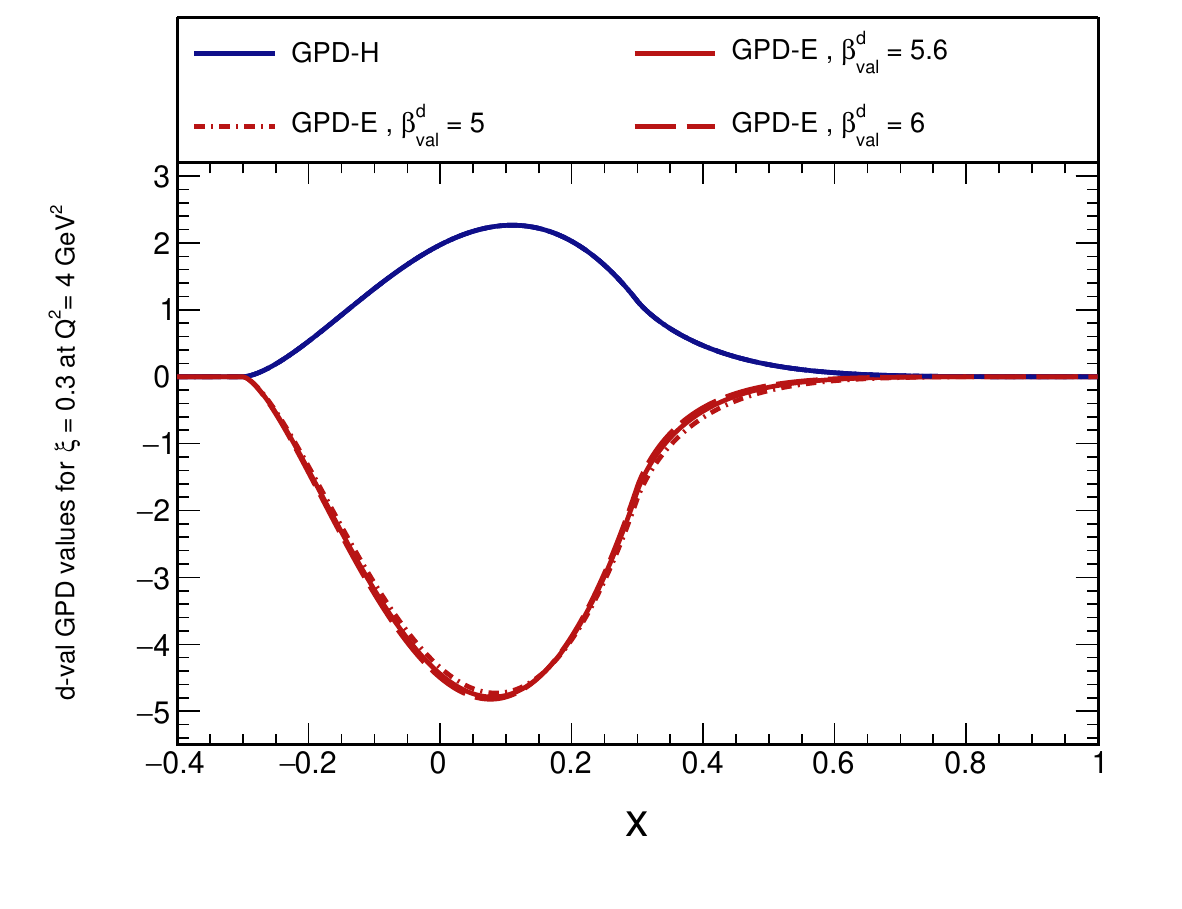} \hspace{1em}
    \includegraphics[width=0.43\linewidth]{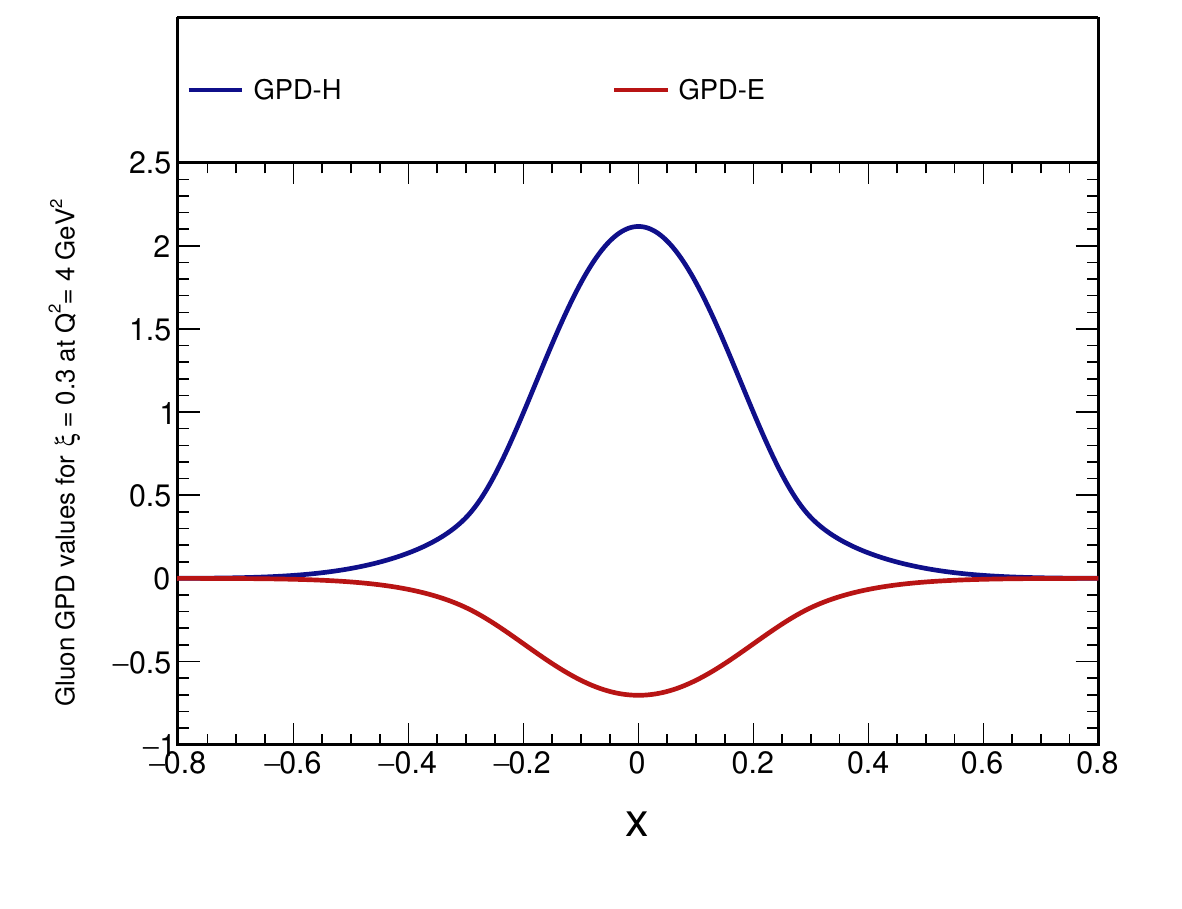} \hspace{1em}
    \includegraphics[width=0.43\linewidth]{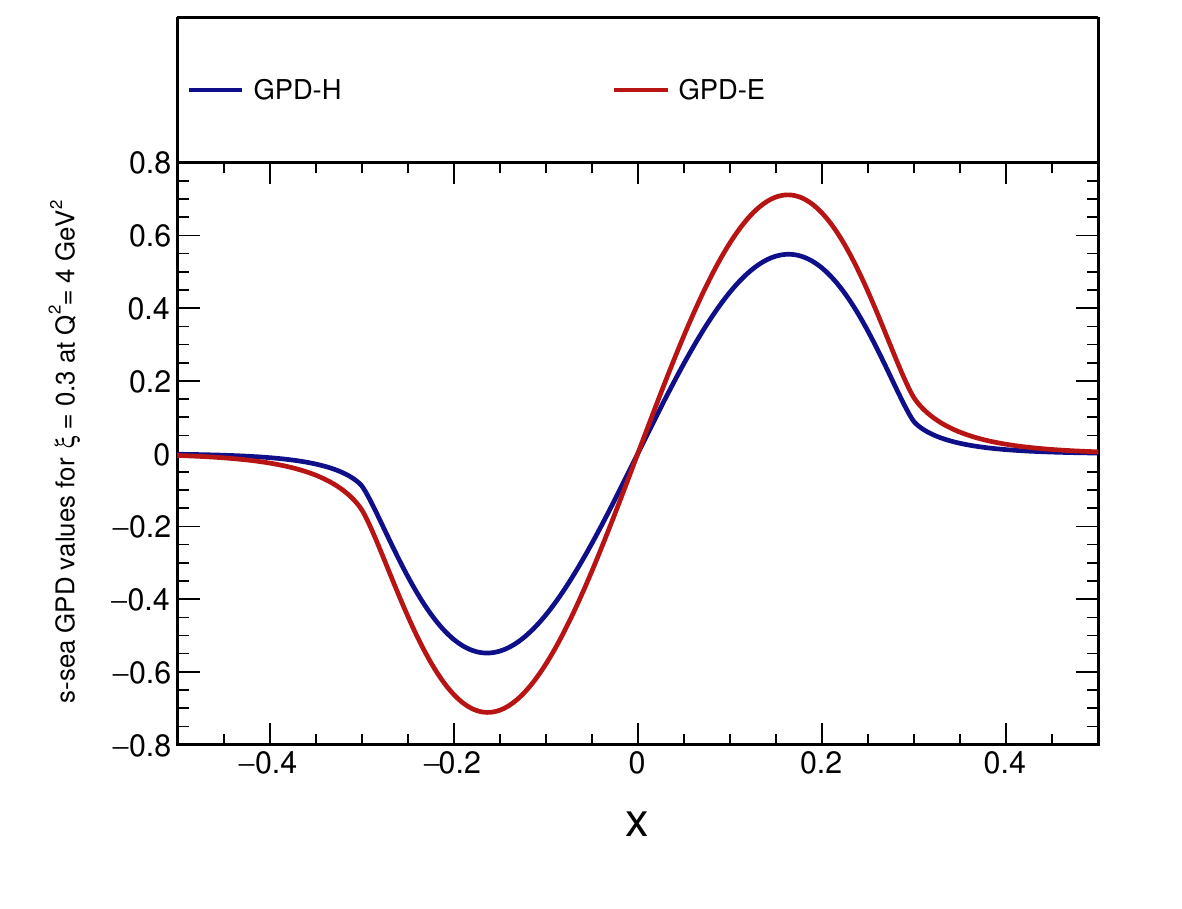}
    \caption{The figures show the H and E-type GPDs used in this work at $\Delta^2=0$ as a function of the momentum fraction $x$ for $\xi=0.3$ and at scale $\mu^2=4\,\rm GeV^2$. The top-left and top-right panels show the $u$ and $d$-valence GPDs, respectively. The bottom-left panel shows the gluon GPDs and the bottom-right panel shows the $s$-sea quark GPDs. The solid blue curves show the GPD-$H$, the solid red curves show the GPD-$E$. In the upper panel the solid red curves correspond to the first variant of the $\beta_{\rm val}^{q=u,d}$, the dash-dotted red curves show the second variant, while the dashed red curves show the third variant corresponding to Table. \ref{tab}.}
    \label{fig:x2}
\end{figure*}
The $E$-type GPDs can be modeled in an analogous fashion to the $H$-type GPDs. Unlike the case of the spin non-flip parton densities, however, their forward limits are not known a priori. Some information can be obtained from the Pauli form factors of the nucleon \cite{Diehl:2004cx,Goloskokov:2008ib}. 

For the valence case we use the parametrization
\begin{eqnarray}
    f^{q_{\rm val}}_e\left(\beta\right)&=&\kappa^q\frac{\Gamma\left(2-\alpha_{\rm val}(0)+\beta^q_{\rm val}\right)}{\Gamma\left(1-\alpha_{\rm val}(0)\right)\Gamma\left(1+\beta^q_{\rm val}\right)}\nonumber\\&\times&\beta^{-\alpha_{\rm val}(0)}\left(1-\beta\right)^{\beta_{\rm val}^q}\theta\left(\beta\right)\,,
    \label{feval}
\end{eqnarray}
where $\alpha_{\rm val}(0)=0.48$, $q=u,d$ and $\kappa^q$ is the contribution of the quarks of flavor $q$ to the anomalous magnetic moment of the proton and satisfies the normalization condition
\begin{equation}
\kappa^{q=u,d}=\int_{-1}^1 dx\, f^{i={\rm val}}_e\left(x\right)\,,
\end{equation}
with 
\begin{eqnarray}
\kappa^{u} &=& 2\kappa^{p} + \kappa^{n} = 1.673 \,, \nonumber\\
\kappa^{d} &=& \kappa^{p} + 2\kappa^{n} = -2.033\, .
\end{eqnarray}
where we have used $\kappa^p=1.793$ and $\kappa^n=-1.913$. 
For the numerical values of $\beta^{q=u,d}_{\rm val}$ we consider three variants
\begin{table}[h]
    \centering
    \begin{tabular}{c c c}
        \hline
        \textbf{Variant} & $\boldsymbol{\beta_{\rm val}^{u}}$ & $\boldsymbol{\beta_{\rm val}^{d}}$ \\
        \hline
         1 & 4  & 5.6 \\
         2 & 10 & 5   \\
         3 & 6  & 6   \\
        \hline
    \end{tabular}
    \caption{Values of the valence quark parameters for the three variants \cite{Goloskokov:2008ib}.}
    \label{tab}
\end{table}
corresponding to the parametrization model proposed by Goloskokov and Kroll (GK model) ~\cite{Diehl:2004cx,Goloskokov:2008ib} at the scale $\mu^2=4\,{\rm GeV}^2$ detailed in Table. \ref{tab}. We have used the first variant in all of our results since it corresponds to the case where all the numerical parameters in the GK parametrization of the GPD-$E$ modeling are non-zero, as can be checked in the table provided in \cite{Goloskokov:2008ib}. However, we also show the model dependence of the valence GPD-$E$ parametrization by considering the remaining two  variants for EIC kinematics. 

Coming to the case of the sea quarks and gluons, a similar parameterization can be constructed as \cite{Goloskokov:2008ib},
\begin{eqnarray}
    f^{g}_e\left(\beta\right) &=& N_{g}\, |\beta|^{-\delta}\left(1-|\beta|\right)^{\beta^{ g}}\,,\nonumber\\f^{s}_e\left(\beta\right) &=& N_{s}\, |\beta|^{-1-\delta}\left(1-|\beta|\right)^{\beta^{ s}}\sign \left(\beta\right)\,.
    \label{glusea}
\end{eqnarray}
In our gluon convention the forward limit of a gluon GPD contains an additional factor of $x$. Therefore the first line of Eq.~\eqref{glusea} is equivalent to the parametrization
$e^g(x)=N_g x^{-1-\delta}(1-x)^{\beta^g}$
used in~\cite{Goloskokov:2008ib}. At $\mu^2=4\,{\rm GeV}^2$ we use $\delta=0.1$ and for the remaining parameters in Eq.~\eqref{glusea}, we take variant 2 of   ~\cite{Goloskokov:2008ib},
\begin{equation}
N_g=-0.873\,,\qquad
N_s=0.155\,,\qquad
\beta^g=6\,,\qquad
\beta^s=7\,.
\label{variant2}
\end{equation}
For simplicity, the sea is assumed to be flavor symmetric. It should be noted that the parameters in Eq.~\eqref{variant2} specify the forward-limit functions only. The $\Delta^2$ dependence used in the present calculation is given separately below.

For the $\Delta$-dependence, we use for sea quarks and gluons, both for $H$ and $E$, a parametrization
${\cal F}(\Delta^2) = \exp(B \Delta^2/2)$, with a common slope parameter $B = 4 ~\rm{GeV}^{-2}$.
For valence $H$-type GPDs, we follow \cite{Belitsky:2001ns}, by using 
\begin{eqnarray}
    \mathcal{F}_{h,\rm val}^u(\Delta^2)&=& \frac{1}{2}F_1^u(\Delta^2)= F_1^p(\Delta^2) + \frac{1}{2} F_1^n(\Delta^2)\,,\nonumber\\\mathcal{F}_{h,\rm val}^d(\Delta^2)&=&F_1^d(\Delta^2)=F_1^p(\Delta^2) + 2 F_1^n(\Delta^2)\,.
\end{eqnarray}
\begin{figure*}[t]
    \centering   
    \includegraphics[width=0.43\linewidth]{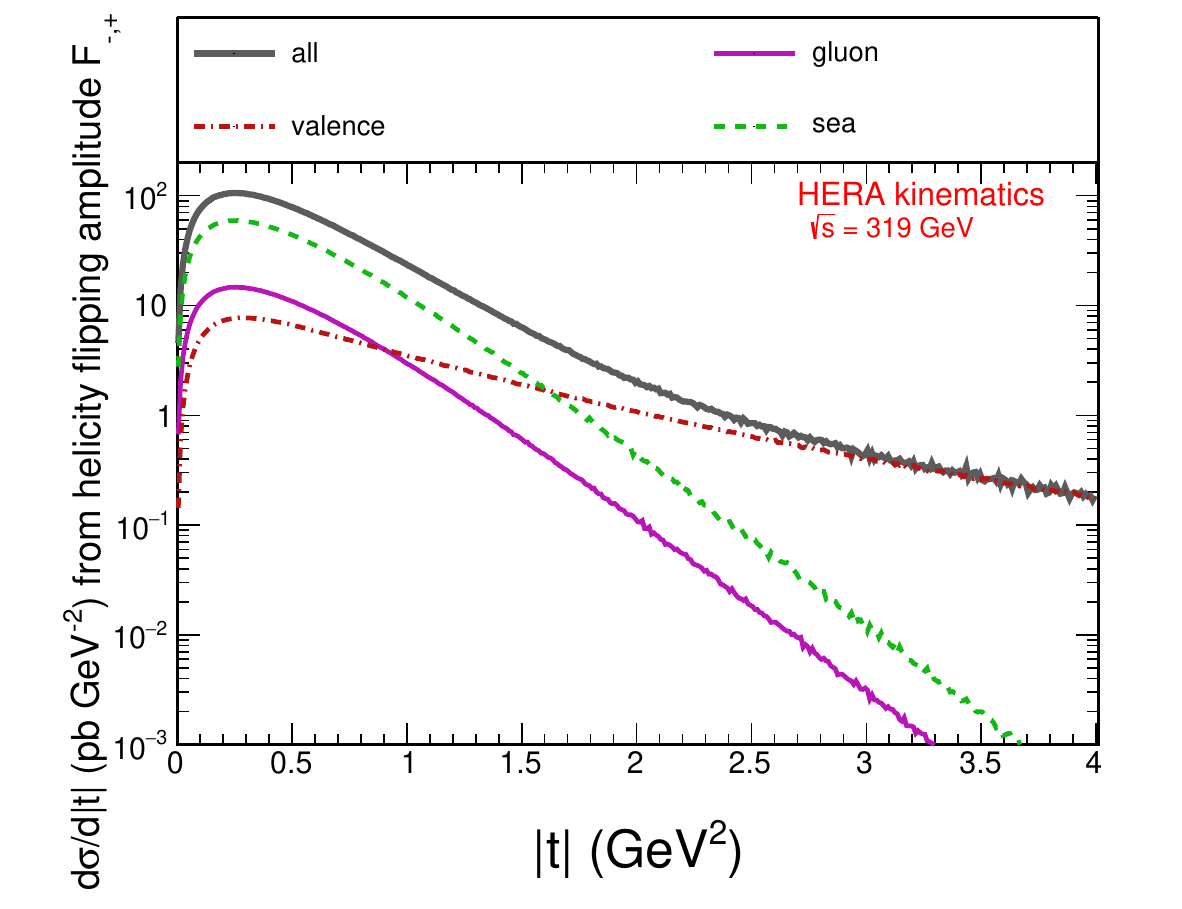}\hspace{1em}\includegraphics[width=0.43\linewidth]{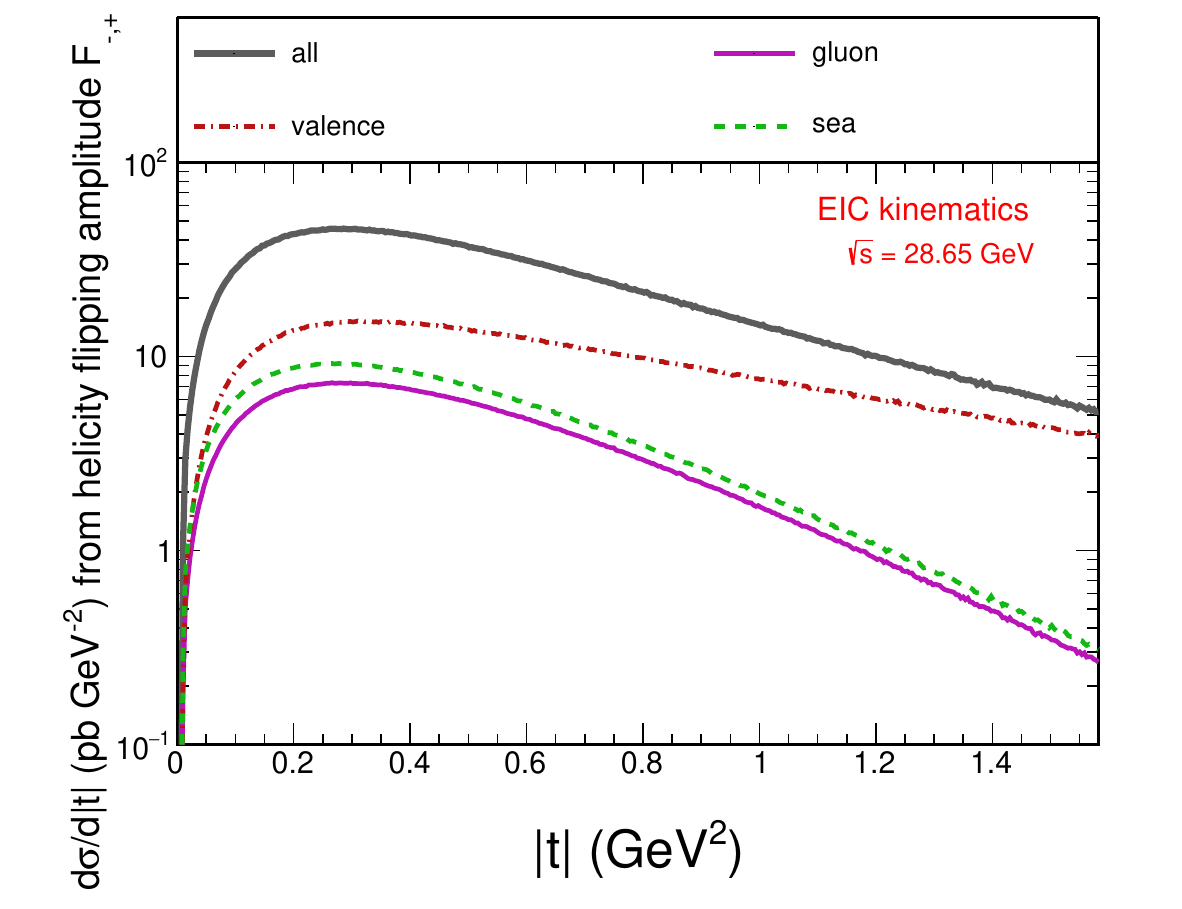}
    \caption{The figures show the differential cross section in the modulus of $t$-Mandelstam variable obtained from the helicity flipping amplitude $F_{-,+}$. The solid black curve denotes the total light-parton contribution, including gluon and sea- and valence-quark exchanges. The solid magenta curve shows the gluon contribution, the red dash-dotted curve the valence-quark contribution, and the green short-dashed curve the sea-quark contribution. The left panel corresponds to the HERA c.m. energy $\sqrt{s}=319\,{\rm GeV}$ without experimental cuts, whereas the right panel is shown for the lowest EIC c.m. energy configuration $\sqrt{s}=28.65\,{\rm GeV}$ with the minimal EIC cuts as specified in the text.}
    \label{fig:t}
 \end{figure*}
 \begin{figure*}[htb]
    \centering   
    \includegraphics[width=0.43\linewidth]{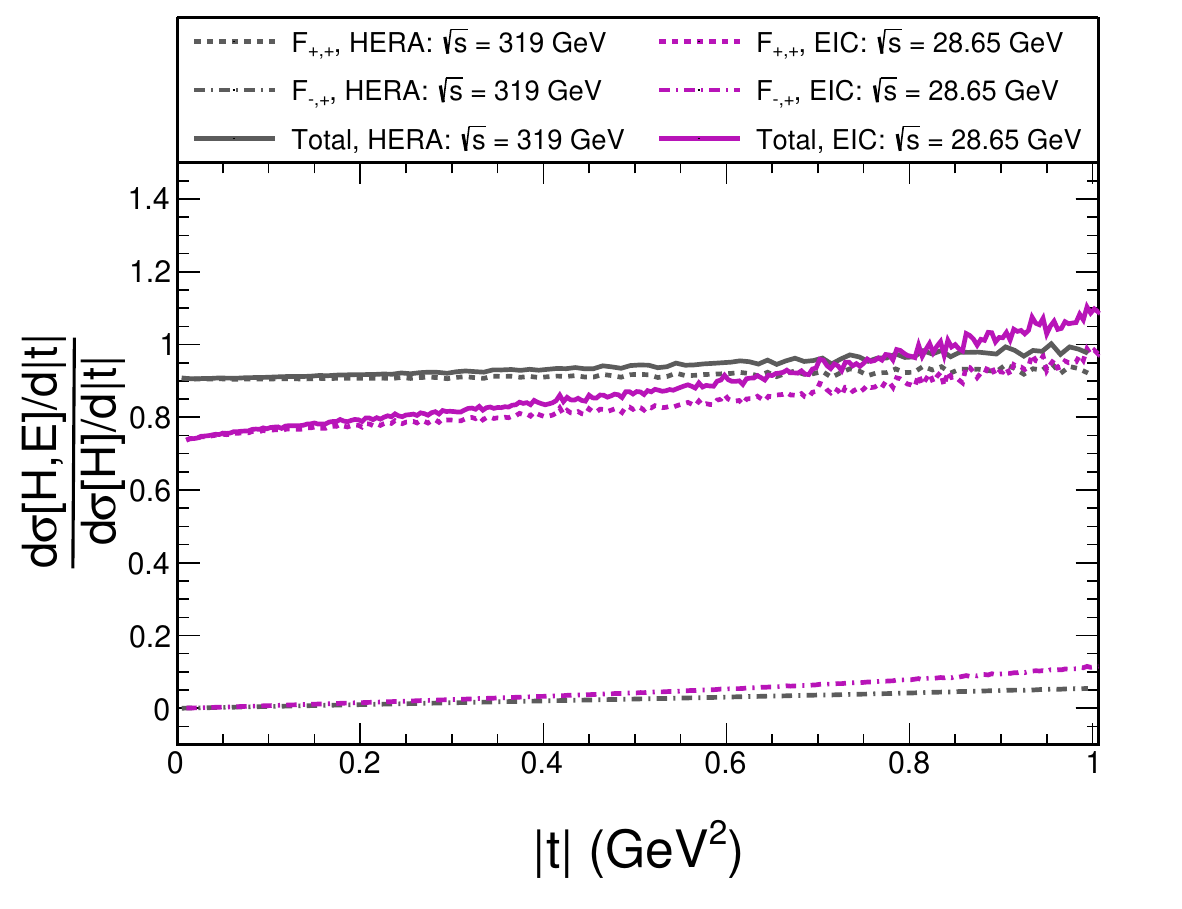}\hspace{1em}\includegraphics[width=0.43\linewidth]{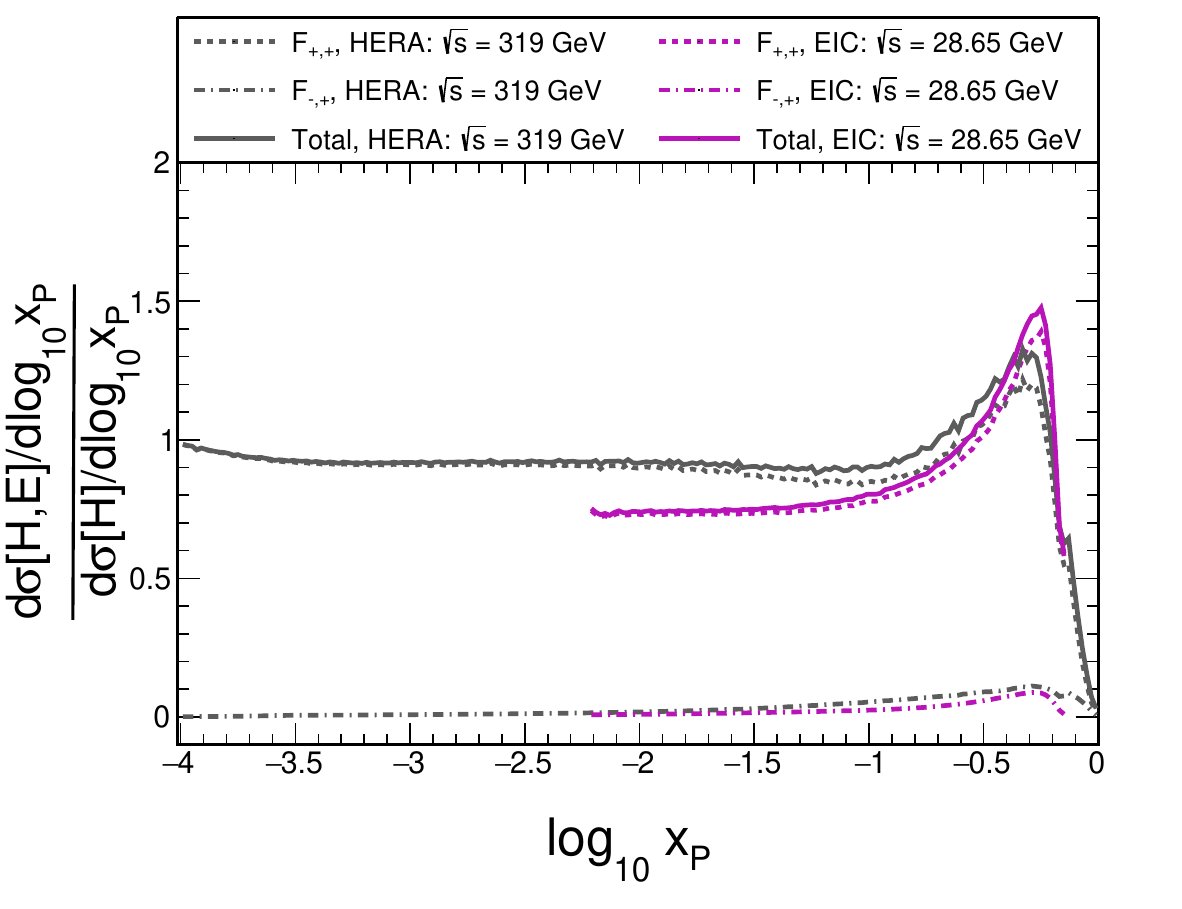}
    \caption{The figures show the ratio of the differential cross section in Mandelstam-$t$ (left panel) and $\log_{10}{x_{\mathbb P}}$ (right panel) obtained from the helicity non-flipping amplitude $F_{+,+}$ with both the GPD-$H$ and GPD-$E$ contributions to that obtained with only the GPD-$H$ contribution. The solid black curve represents this ratio for HERA kinematics with c.m. energy $\sqrt{s}=319\,{\rm GeV}$, whereas the dash-dotted magenta curve represents the ratio for EIC kinematics with c.m. energy $\sqrt{s}=28.65\,{\rm GeV}$  with the minimal EIC cuts as specified in the text. Both the ratios are evaluated for the total light-parton contribution, including gluon, light sea and valence-quark exchanges.}
    \label{fig:ratio}
 \end{figure*}
The factor $1/2$ in the u-valence case reflects the normalization $\int_0^1 u_{\rm val}(x)\,dx=2$.

For the $E$-type GPDs the corresponding factors have to preserve the normalization $\int_0^1 dx\,E_{\rm val}^q(x,0,t)=F_2^q(t)$.
Since the forward functions in Eq.~\eqref{feval} are already normalized to $\kappa^u$ and $\kappa^d$, we use the normalized form factors
\begin{eqnarray}
    \mathcal{F}_{e,\rm val}^u(\Delta^2)&=& \frac{F_2^u(\Delta^2)}{\kappa^u}=\frac{2 F_2^p(\Delta^2) + F_2^n(\Delta^2)}{\kappa^u}\,,\nonumber\\\mathcal{F}_{e,\rm val}^d(\Delta^2)&=&\frac{F_2^d(\Delta^2)}{\kappa^d}=\frac{F_2^p(\Delta^2) + 2 F_2^n(\Delta^2)}{\kappa^d}\,.
\end{eqnarray}
With these definitions ${\cal F}_{e,{\rm val}}^{u,d}(0)=1$.
Neglecting strange form factors, the flavor form factors can be expressed in terms of the proton and neutron form factors as,
\begin{eqnarray}
F_{1,2}^u(t)&=& 2F_{1,2}^p(t)+F_{1,2}^n(t)\,, \nonumber\\ F_{1,2}^d(t) &=& F_{1,2}^p(t)+2F_{1,2}^n(t)\,.
\label{flavorFF}
\end{eqnarray}
Now these form factors can be expressed in terms of the Sachs electric and magnetic form factors for the proton and the neutron as,
\begin{eqnarray}
    F_1^{p,n}(\Delta^2)&=&\frac{G_E^{p,n}(\Delta^2)+\tau_{p,n}G_M^{p,n}(\Delta^2)}{1+\tau_{p,n}}\,,\nonumber\\F_2^{p,n}(\Delta^2)&=&\frac{G_M^{p,n}(\Delta^2)-G_E^{p,n}(\Delta^2)}{1+\tau_{p,n}}\,,
\end{eqnarray}
where $\tau_{p,n}=\frac{|\Delta^2|}{4m_{p,n}^2}$. The ratios of the Sachs electric to magnetic form factors is parametrized as \cite{Punjabi:2015bba},
\begin{eqnarray}
    \mu_n\frac{G^n_E}{G^n_M}&=&\frac{A_1\tau_n}{1+A_2\sqrt{\tau_n}+A_3\tau_n}\,,\nonumber\\\mu_p\frac{G^p_E}{G^p_M}&=&\frac{1+B_0\tau_p+B_1\tau_p^2+B_2\tau_p^3}{1+B_3\tau_p+B_4\tau_p^2+B_5\tau_p^3+B_6\tau_p^4}\,,
    \label{ratio}
\end{eqnarray}
where $A_1=2.6316,\,A_2=4.118,\,A_3=0.29516$ and $B_0=-5.7891,\,B_1=14.493,\,B_2=-3.5032,\,B_3=-5.5839,\,B_4=12.909,\,B_5=0.88996,\,B_6=1.542$. The prefactors to the aforementioned ratios in Eq.~\eqref{ratio} are $\mu_n=-1.913$ and $\mu_p=2.793$. 
For the Sachs magnetic form factors entering the ratios in Eq.~\eqref{ratio}, the Kelly fit \cite{Kelly:2004hm} is used for $G_M^{p,n}$,
\begin{eqnarray}
    G^n_M &=&\mu_n\left(\frac{1+a_0\tau_n}{1+a_1\tau_n+a_2\tau_n^2+a_3\tau_n^3}\right)\,,\nonumber\\G^p_M &=& \mu_p \left(\frac{1+b_0\tau_p}{1+b_1\tau_p+b_2\tau_p^2+b_3\tau_p^3}\right)\,,
\end{eqnarray}
where $a_0=2.33,\,a_1=14.72,\,a_2=24.2,\,a_3=84.1$ and $b_0=0.12,\,b_1=10.97,\,b_2=18.86,\,b_3=6.55$. 
The so--constructed quark-GPDs fulfill the sum--rules
\begin{eqnarray}
\sum_q e_q \int_{-1}^1 dx \, H^q_{\rm val}(x,\xi,\Delta^2) &=& F_1^{p}(\Delta^2) \,,\nonumber\\ \sum_q e_q \int_{-1}^1 dx \, E^q_{\rm val}(x,\xi,\Delta^2) &=& F_2^{p}(\Delta^2).    
\end{eqnarray}

In Fig.~\ref{fig:x1} and Fig.~\ref{fig:x2} we show GPD-$E$ together with GPD-$H$ for u-valence, d-valence, gluon and s-sea quark exchanges as a function of $x$. For the valence quark E-type GPDs, we show for all the three possible variants of the numerical parameter $\beta_{\rm val}^{q=u,d}$ (see Table. \ref{tab}) considered in the GK parametrization. Fig.~\ref{fig:x1} is for rather small $\xi= 0.01$ and Fig.~\ref{fig:x2} is for $\xi= 0.3$.
In both cases the $E$-type GPDs are of similar size as the $H$-type which strongly motivates the analysis of the role of $E$-GPDs in the cross section for dijet electroproduction.

\section{Results and Discussions}
\label{sec:4}
In this section we discuss separately the partonic composition of the proton helicity-flip contribution and the relative size of the helicity-conserving and helicity-flip amplitudes. As in~\cite{Chall:2026oes}, we use the HERA center-of-mass energy $\sqrt{s}=319\,{\rm GeV}$ as a reference and compare it with the EIC energy configurations considered below.

We first consider the cross section obtained from the proton helicity-flip amplitude $F_{-,+}$ alone. Figure~\ref{fig:t} shows the distribution in the absolute value of the momentum transfer, $|t|$, separated into gluon, sea-quark and valence-quark contributions. The left panel corresponds to the HERA energy scale, $\sqrt{s}=319\,{\rm GeV}$, whereas the right panel is shown for the lowest EIC energy configuration, $\sqrt{s}=28.65\,{\rm GeV}$.

To explore an extensive phase space at the HERA energy scale, we take the same broad kinematic region as in   ~\cite{Chall:2026oes}, without imposing the experimental HERA cuts:
$4<Q^2<10^4\,{\rm GeV}^2$, $0<y<1$, $0.01<z<0.99$,
$1<p_\perp<15\,{\rm GeV}$ and $0<\Delta_\perp<3\,{\rm GeV}$.

We see that in both panels the helicity-flip contribution is dominated by valence-quark exchange. The sea-quark and gluon terms are visible at small $|t|$, but they decrease much faster with increasing momentum transfer. The valence contribution has a substantially slower $|t|$ falloff and therefore practically determines the total $F_{-,+}$ cross section at larger $|t|$. This behavior is especially pronounced in the HERA-energy panel, where the total and valence curves become almost indistinguishable over a large part of the plotted range. We have also checked that the interference effects coming from the interplay between the contribution of each partonic component to the helicity flippping amplitude $F_{-,+}$ are in practice negligibly small, therefore we won't divulge into further discussions on the uncertainties from the subleading gluon and quark contributions to $F_{-,+}$.

The rise of the helicity-flip distribution away from the minimum momentum transfer is also consistent with the kinematic structure of Eq.~\eqref{str}. The amplitude $F_{-,+}$ contains the transverse momentum-transfer factor multiplying $E^{g,q}$ and must vanish at the minimal momentum transfer. The distribution therefore reaches its maximum at a nonzero value of $|t|$ before decreasing at larger momentum transfer.
 
\begin{figure*}[t]
    \centering 
    \includegraphics[width=0.43\linewidth]{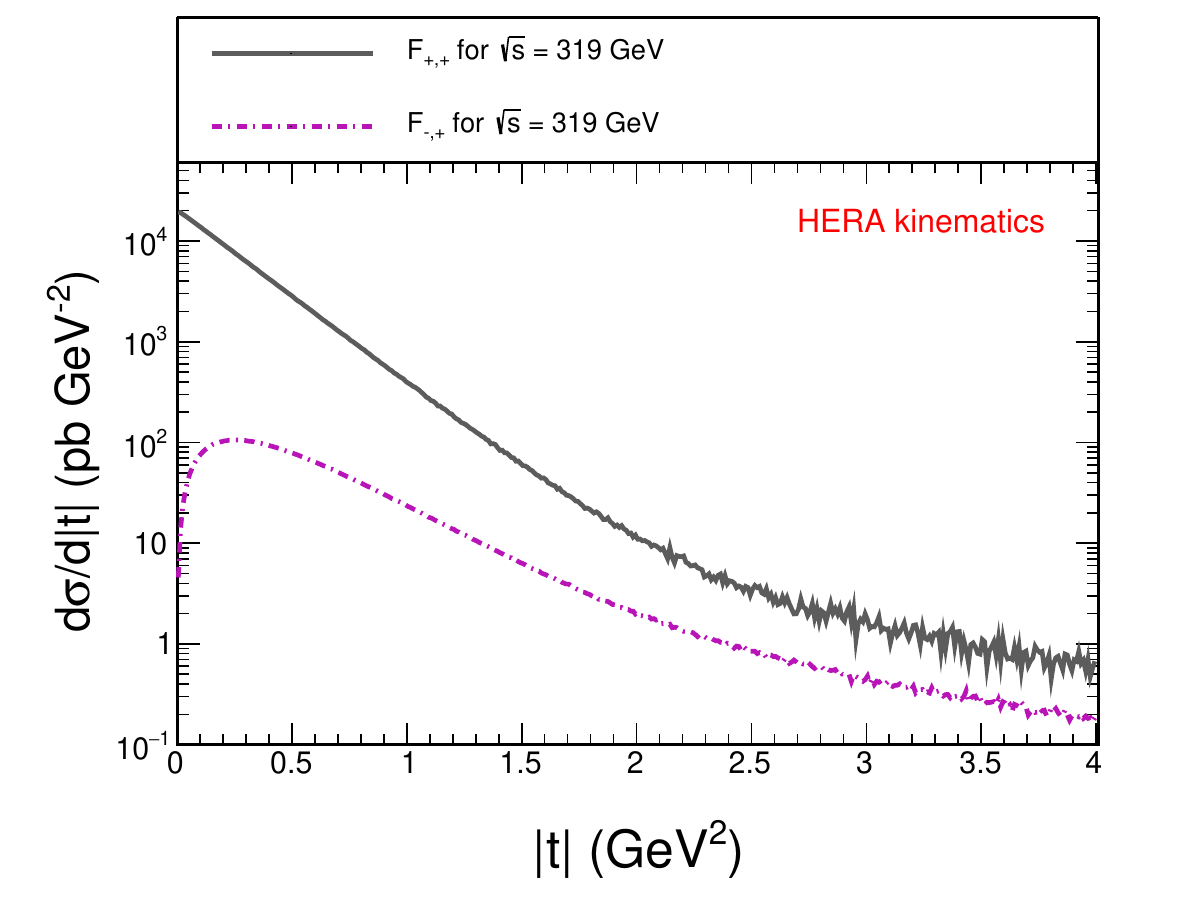}\hspace{1em}
    \includegraphics[width=0.43\linewidth]{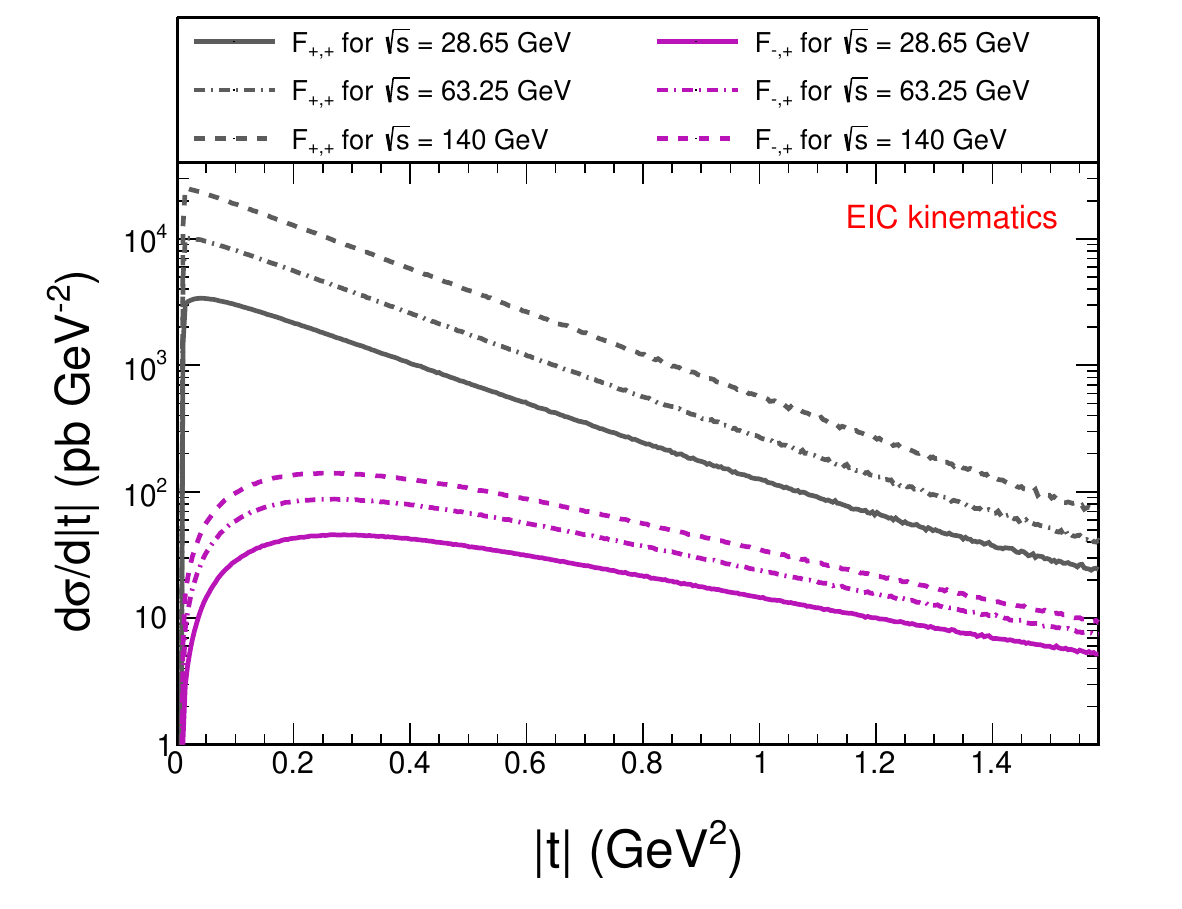}\hspace{1em}
    \includegraphics[width=0.43\linewidth]{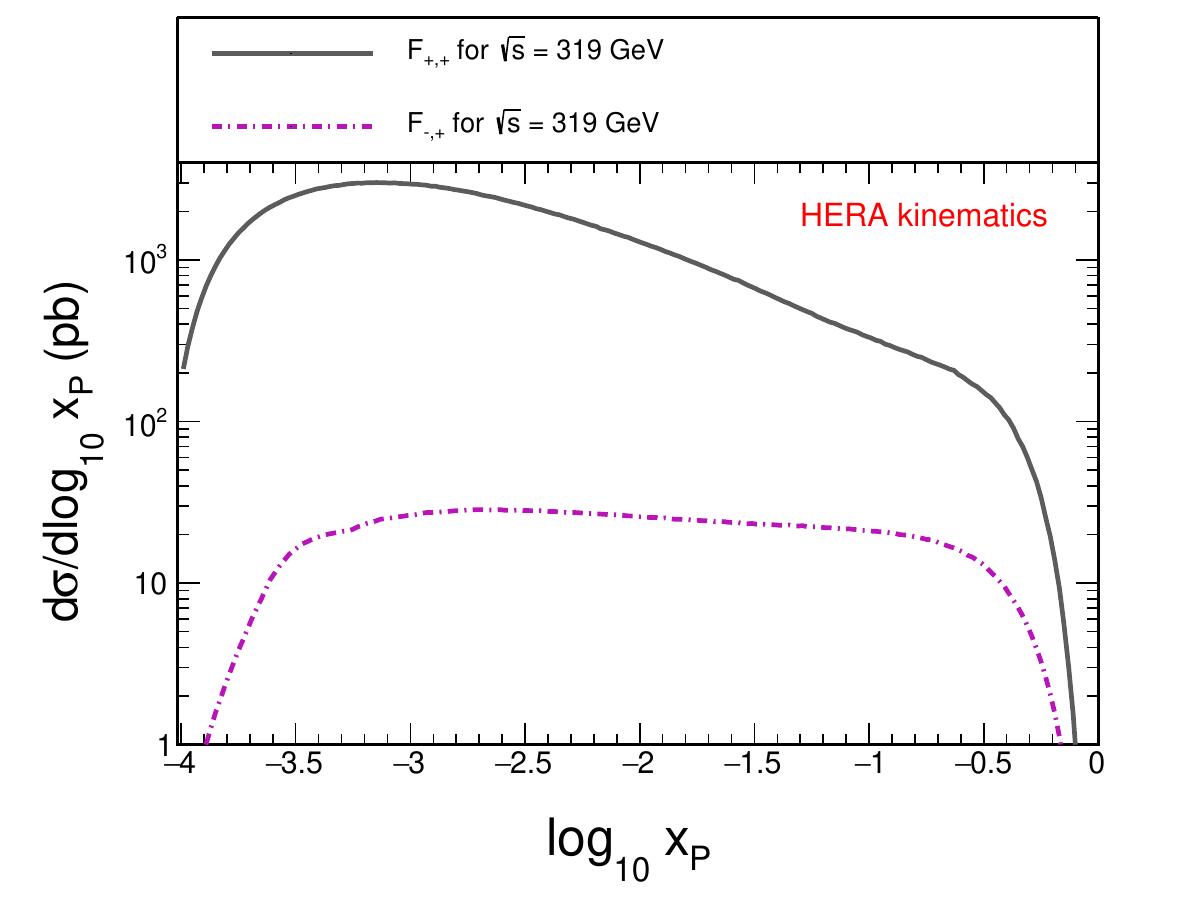}\hspace{1em}\includegraphics[width=0.43\linewidth]{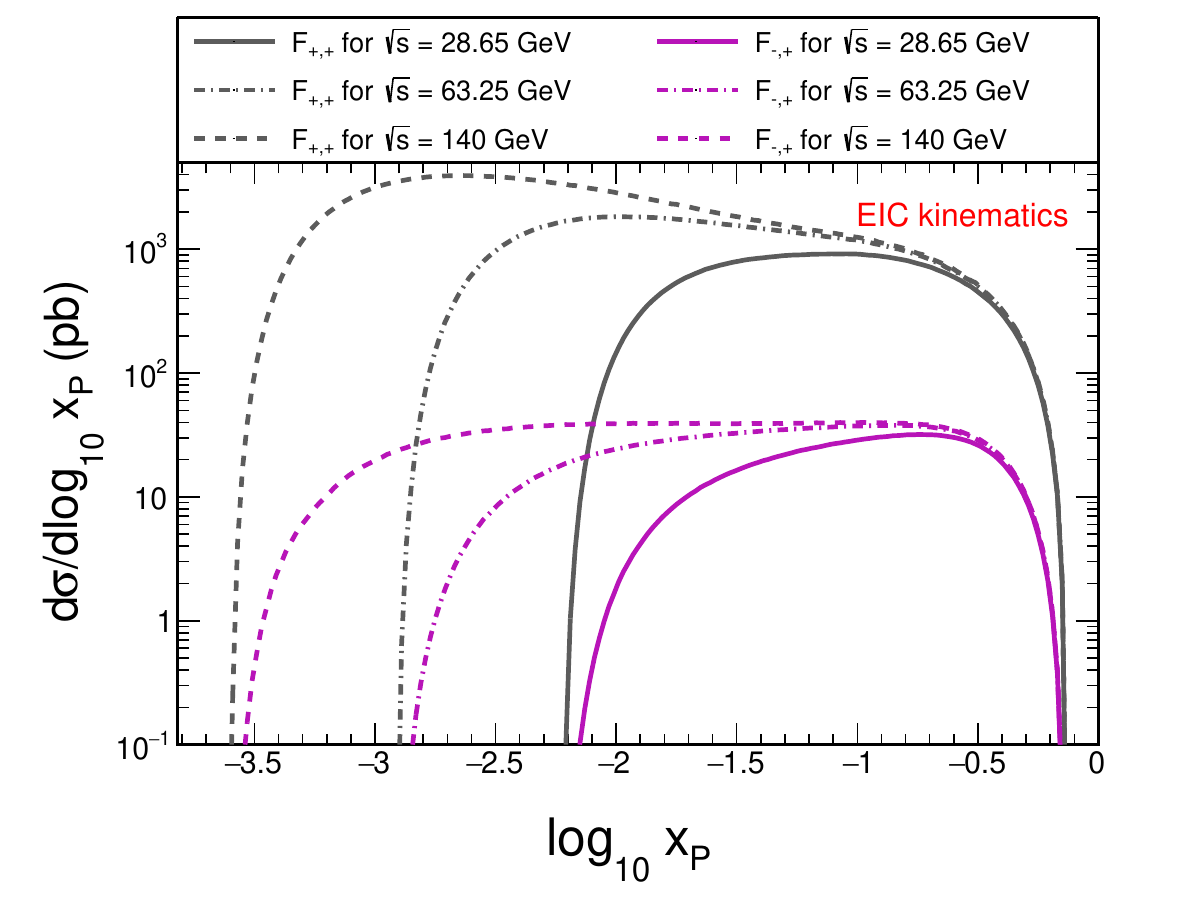}
    \caption{Comparison of the proton helicity-conserving and helicity-flip contributions to exclusive dijet electroproduction. The upper panels show the differential cross section in $|t|$, while the lower panels show the distribution in $\log_{10}x_{\mathbb P}$. The black curves denote the contribution from the helicity-conserving amplitude $F_{+,+}$ and the magenta curves show the contribution from the helicity-flipping amplitude $F_{-,+}$. The left panels correspond to the HERA c.m. energy $\sqrt{s}=319\,{\rm GeV}$ with minimal cuts in an extended phase space. The right panels show the distributions for the EIC configurations with c.m. energies $\sqrt{s}=28.65\,{\rm GeV}$ (solid), $\sqrt{s}=63.25\,{\rm GeV}$ (dash-dotted) and $\sqrt{s}=140\,{\rm GeV}$ (dashed) with the minimal EIC cuts specified in the text. All curves include gluon, light sea and valence-quark exchanges.}
    \label{fig:xp1}
 \end{figure*} 

We now compare the two proton helicity components. For the EIC we take the three beam-energy configurations
\begin{eqnarray}
(E_e,E_p)&=&(5,41)\ {\rm GeV}\,,
\nonumber\\
(E_e,E_p)&=&(10,100)\ {\rm GeV}\,,
\nonumber\\
(E_e,E_p)&=&(18,275)\ {\rm GeV}\,,
\end{eqnarray}
which correspond approximately to $\sqrt{s}=28.65$, $63.25$ and $140\,{\rm GeV}$, respectively. We impose the relevant minimal cuts for the EIC,
\begin{equation}
10^{-4}<y<0.95,\qquad 10^{-5}<x_{\rm Bj}<0.95,
\end{equation}
together with 
\begin{eqnarray}
&&0.7<Q^2<1000\,{\rm GeV}^2,\quad
0.01<|t|<1.6\,{\rm GeV}^2,\\&& p_{\perp}\geq 1\,{\rm GeV}\,,\qquad 0.01<z<0.99\,.
\label{minimal}
\end{eqnarray}

In order to summarize our results on the role of GPD $E$ 
in Fig.~\ref{fig:ratio} we show the ratio of the cross section for dijet production when including both $H$ and $E$ GPDs to the cross section when only GPD-$H$ is included as a function of Mandelstam-$|t|$ (left panel) and $x_{\mathbb P}$ (right panel).
We show separately the helicity conserving (dotted line) and helicity flipping (dash-dotted line) components as well as the one from the total contribution (solid line).
The dotted curves show the ratio of the obtained cross-section distribution from the helicity conserving amplitude $F_{+,+}$ with both the GPD-$H$ and GPD-$E$ contributions to the cross section obtained with only the GPD-$H$ contribution. For the case of HERA (shown by black curves), as expected we see that the addition of second term containing the GPD-$E$ in the first line of Eq.~\eqref{str} has almost no effect on this ratio in the HERA kinematical region of lower $x_{\mathbb P}$, but it shows a bump at relatively high $x_{\mathbb P}$ in the right panel of Fig.~\ref{fig:ratio}. For the EIC kinematics, we can clearly see that the effect of GPD-$E$ is crucial and at larger $|t|$ and $x_{\mathbb P}$ it overtakes the HERA curves shown in black emerging from the enhanced GPD-$E$ contribution at large $\xi$ or $x_{\mathbb P}$.
At the EIC, we observe a decrease at small $x_{\mathbb P}$ stemming from the negative interference effect of $H$ and $E$-GPD terms and an enhancement at large $x_{\mathbb P}$ ($\log_{10}x_{\mathbb P} \lesssim  -0.5$) due to the factor $\frac{\xi^2}{1 - \xi^2}$ in front of $E^{g,q}$ in Eq.~\eqref{str}.

The dash-dotted curves in Fig.~\ref{fig:ratio} show the ratio of the cross-section distribution from the helicity flipping amplitude $F_{-,+}$ (which in essence is a manifestation of only GPD-$E$) to the cross section obtained with only the GPD-$H$ contribution. This ratio is extremely small for almost the entire range of $|t|$ and $x_{\mathbb P}$ for HERA kinematics and for EIC kinematics except at high $x_{\mathbb P}$ and $|t|$ where we see a bump in the ratio.

\begin{figure*}[t]
    \centering   
    \includegraphics[width=0.43\linewidth]{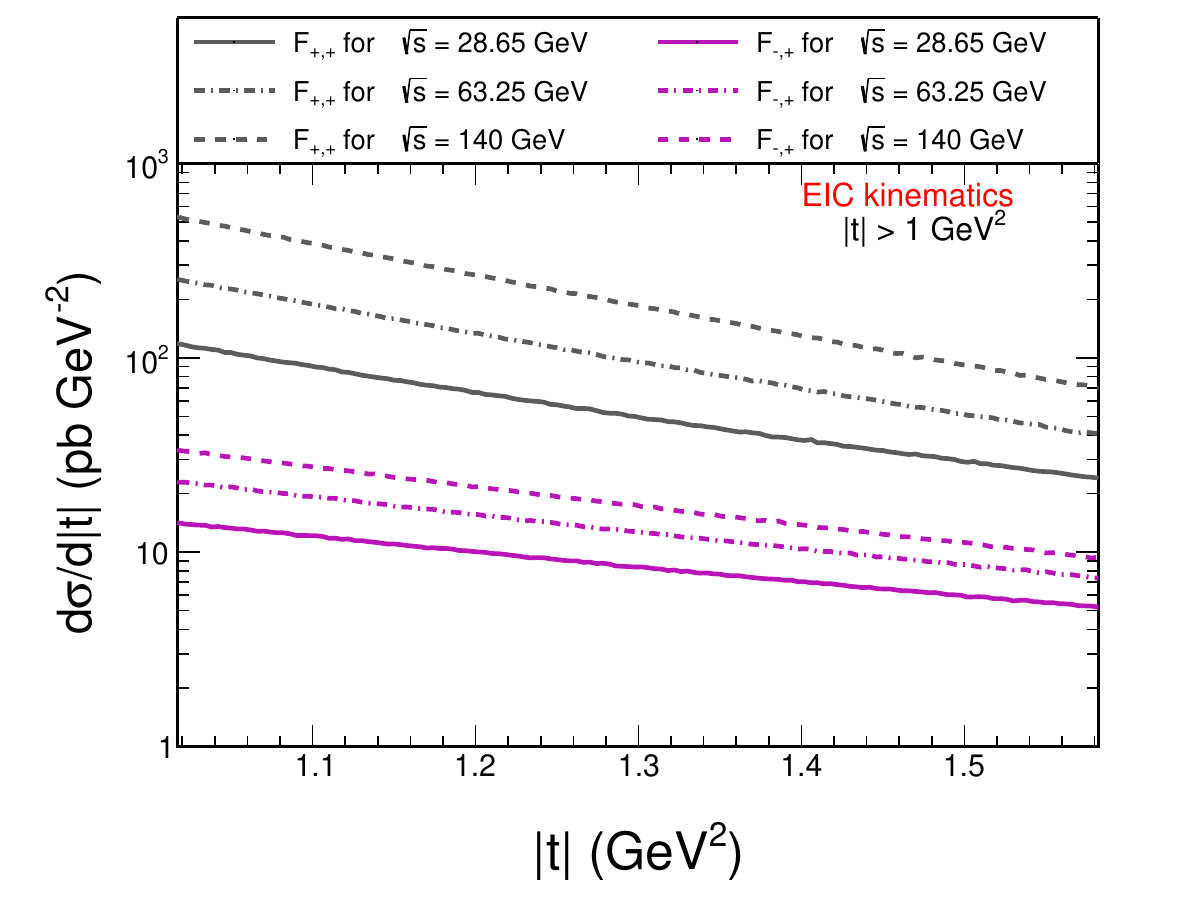}\hspace{1em}
    \includegraphics[width=0.43\linewidth]{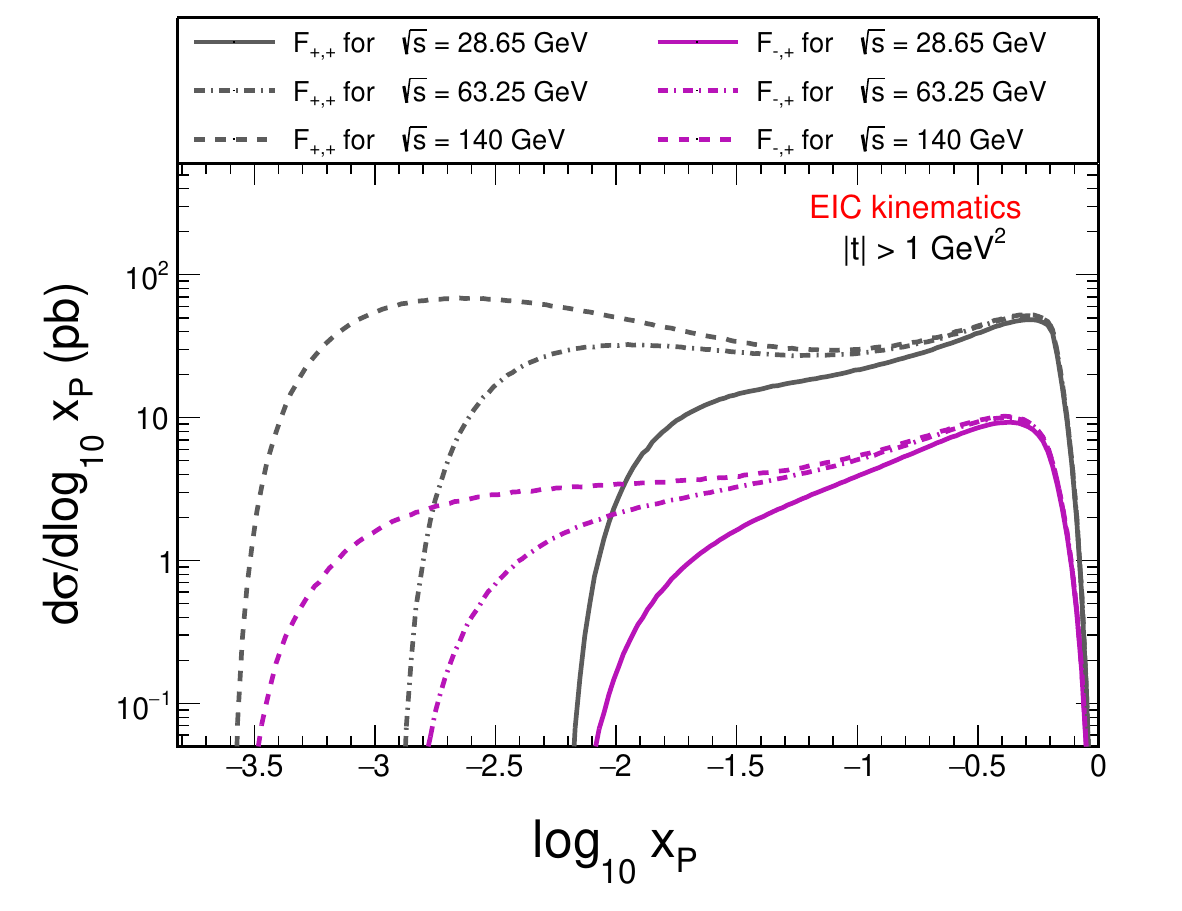}
    \caption{The figures show the differential cross section in Mandelstam-$|t|$ (left panel) and $\log_{10}x_{\mathbb{P}}$ (right panel)
    obtained from the both the helicity non-flipping amplitude $F_{+,+}$ (in black) and from the flipping amplitude $F_{-,+}$ (in magenta) for EIC configurations with c.m. energies $\sqrt{s}=28.65\,{\rm GeV}$ (solid), $\sqrt{s}=63.25\,{\rm GeV}$ (dash-dotted) and $\sqrt{s}=140\,{\rm GeV}$ (dashed) with the minimal EIC cuts specified in the text except for the cut $|t|> 1\, {\rm GeV^2}$. All curves include gluon and light sea- and valence-quark exchanges which are dominant for EIC kinematics.}
    \label{fig:newEIC}
 \end{figure*}
\begin{figure*}[htb]
    \centering   
    \includegraphics[width=0.43\linewidth]{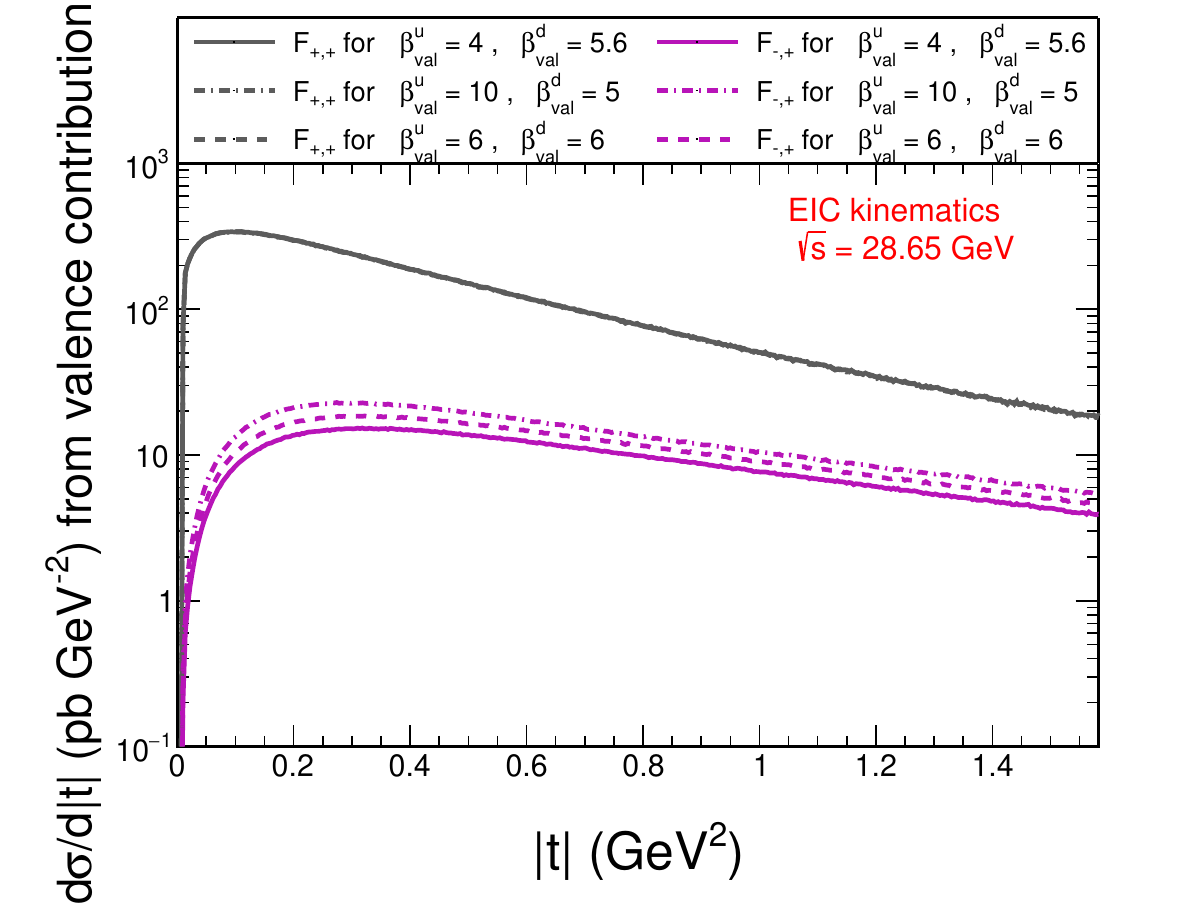}\hspace{1em}
    \includegraphics[width=0.43\linewidth]{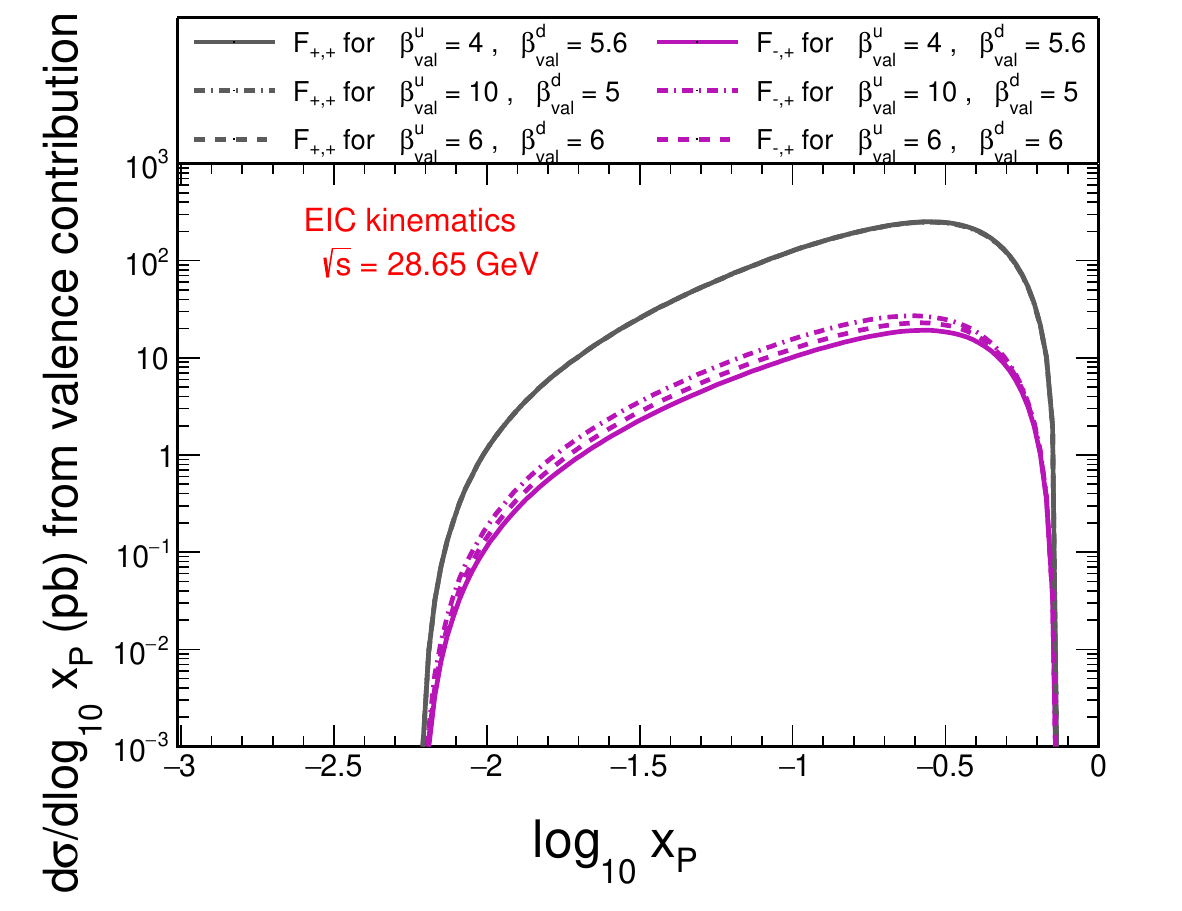}
    \caption{The figures show the differential cross section in Mandelstam-$|t|$ (left panel) and $\log_{10}x_{\mathbb{P}}$ (right panel) obtained from both the helicity non-flipping amplitude $F_{+,+}$ (in black) and from the flipping amplitude $F_{-,+}$ (in magenta) for EIC configuration with c.m. energy $\sqrt{s}=28.65\,{\rm GeV}$ with the the minimal EIC cuts specified in the text. The solid curves represent the first variant, the dash-dotted curves denote the second variant and the dashed curves denote the third variant of the $\beta^q_{\rm val}$ parameters detailed in Table. \ref{tab}. All curves include only the light valence-quark exchanges which are dominant for EIC kinematics.}
    \label{fig:val}
 \end{figure*}
The upper panels of Fig.~\ref{fig:xp1} show the $|t|$ distributions. At the HERA energy scale the helicity-conserving contribution $F_{+,+}$ is largest close to the forward region and decreases rapidly with $|t|$. The helicity-flip contribution $F_{-,+}$ vanishes toward the minimum momentum transfer, rises to a broad maximum and then decreases more slowly. This contribution is relatively enhanced at larger $|t|$ compared to the helicity conserving contribution. A similar pattern is present for the EIC configurations. The relative importance of $F_{-,+}$ increases with $|t|$, whereas the helicity-conserving curves fall more steeply. The three EIC energies show a visible energy dependence, particularly for the nonflip component. The separation of the two helicity contributions at moderate $|t|$ is also present for the EIC kinematics considered here. 

In the lower panels of Fig.~\ref{fig:xp1} we show the corresponding distributions in $x_{\mathbb P}$. For the HERA-energy reference calculation, the helicity-conserving $F_{+,+}$ contribution is concentrated at small $x_{\mathbb P}$, whereas the obtained helicity-flip contribution from $F_{-,+}$ is shifted toward larger $x_{\mathbb P}$. This behavior follows the valence-quark dominance of the helicity-flipping contribution manifested by the GPD-$E$ in this kinematical region seen in Fig.~\ref{fig:t}. It is also consistent with ~\cite{Chall:2026oes}, where the valence contribution was found to become more important at larger $x_{\mathbb P}$.
For the EIC kinematics similar qualitative separation is seen. The helicity-conserving contribution moves toward smaller $x_{\mathbb P}$ as the center-of-mass energy is increased, whereas the helicity-flip contribution stays concentrated at larger $x_{\mathbb P}$. 


From Figs.~\ref{fig:t} and \ref{fig:xp1} we see a consistent pattern. The helicity-conserving cross section is controlled mainly by the $H$-type GPDs and is largest in the small-$x_{\mathbb P}$, small-$|t|$ region. The helicity-flip contribution directly probes $E^{g,q}$ and, receives a large valence-quark contribution. It therefore becomes relatively more important at larger $x_{\mathbb P}$ and at finite momentum transfer. We also performed the calculation for EIC kinematics using a cut on Mandelstam-$|t|$ as $|t|>1\,\rm GeV^2$ in Fig.~\ref{fig:newEIC} and found that compared to the picture with the minimal EIC cut on $|t|$ (see Eq.~\eqref{minimal}) given in the bottom right panel of Fig.~\ref{fig:xp1}, the cross section distribution becomes relatively more concentrated in the larger $x_{\mathbb P}$ region.

Lastly, in Fig.~\ref{fig:val}, we test the model dependence of the valence parametrization for EIC kinematics in the lowest c.m. configuration $\sqrt{s}=28.65\,\rm GeV$ by considering the remaining last two variants for the choice of $\beta^{q=u,d}_{\rm val}$ (see Table. \ref{tab}) in the GK model and observe the separation of the helicity-flipping valence quark exchange contributions stemming from the three different variants in Table. \ref{tab}. The second variant (shown by magenta dash-dotted curves) leads over the the third variant (shown by magenta dashed curves) which leads over the first variant (shown by magenta solid curves) for both the cross section distributions in Mandelstam-$|t|$ and $x_{\mathbb P}$.

\section{Conclusions}
\label{sec:5}
In this work we have taken into account both the proton helicity conserving as well as flipping contribution associated with the $E$-type GPD. The hard scattering part is the same as in  \cite{Braun:2005rg,Chall:2026oes}, while the proton matrix element now includes both helicity-conserving and helicity-flip amplitudes. The latter is attributed to $E^{g,q}$. We find that the $E$-GPD
contribution has a different partonic composition from the helicity-conserving one. It is dominated by valence-quark exchange over a broad range of $|t|$. The sea and gluon contributions are relevant mainly at small momentum transfer and decrease more rapidly with $|t|$. The valence contribution therefore determines the behavior of the helicity-flip cross section at larger $|t|$.

The major difference is clearly visible in the $x_{\mathbb P}$ distributions. The helicity-conserving contribution is concentrated at relatively smaller $x_{\mathbb P}$, whereas the helicity-flip contribution is shifted toward somewhat larger $x_{\mathbb P}$. For the EIC energy configurations considered here, the two contributions become of the same order of magnitude in the region of large $t$ and $x_{\mathbb P}$. The effect is increased for the lowest EIC energy configuration. 
The valence part of $E$ is constrained by the nucleon Pauli form factors, while the sea and gluon parts are less constrained. 

In this work, for modeling the $E$-type GPDs, we have used the Goloskokov-Kroll (GK) model and have also consequently tested the model dependence of our results especially for the different variants of parameters entering the valence parametrization. In \cite{Goloskokov:2008ib}, Goloskokov and Kroll suggested measuring the target transverse spin asymmetry $A_{U,T}$ for the exclusive electroprodution of vector mesons. At that time the preliminary HERMES data \cite{HERMES:2009gyx} was consistent with the GK model for the $E$-type GPDs. There were further measurements by the HERMES and COMPASS collaborations \cite{HERMES:2015zqh,COMPASS:2016ium}, but the statistics did not allow for precise tests of the GPD-$E$.
The exclusive dijet electroproduction could provide another way for testing it.

The results indicate that the unpolarized exclusive dijet cross section in the EIC kinematical region has a sizable contribution from the $E$-type GPDs, especially at finite $|t|$ and at moderate-to-large
$x_{\mathbb P}$.
The separate helicity components shown in this work are a theoretical decomposition; a direct experimental separation would require polarization-sensitive information. It would be valuable to identify the helicity-flip dijet electroproduction cross section at higher $x_{\mathbb P}$ and $|t|$ experimentally at the EIC. A detailed detector-level feasibility study is beyond the scope of the present analysis.

\begin{acknowledgments}
This work was supported by the Polish National Science Center Grant No.~UMO-2023/49/B/ST2/03665. 
\end{acknowledgments}
\bibliography{dijet.bib}
\end{document}